\documentclass[twocolumn, times, astrosymb]{aastex7}

\usepackage{amsmath,amstext}
\usepackage{graphicx}
\usepackage{booktabs}
\usepackage{hyperref}
\usepackage{caption}  

\newcommand*\erp[1]{{\color{blue} #1}}

\shorttitle{Quantitatively Identifying FU Orionis YSOs}
\shortauthors{Portnoi et al.}

\submitjournal{AAS Journals}

\turnoffediting

\begin{document}

\title{Quantitative Spectroscopic Diagnostics for FU Orionis-Type Young Stellar Objects}

\correspondingauthor{Evan R. Portnoi}

\author{Evan R. Portnoi}
\affiliation{Division of Physics, Mathematics, and Astronomy, California Institute of Technology, Pasadena, CA 91125, USA}
\email{eportnoi@caltech.edu}

\author{Lynne A. Hillenbrand}
\affiliation{Division of Physics, Mathematics, and Astronomy, California Institute of Technology, Pasadena, CA 91125, USA}
\email{lah@astro.caltech.edu}

\author[0000-0002-9540-853X]{Adolfo S. Carvalho}
\affiliation{Division of Physics, Mathematics, and Astronomy, California Institute of Technology, Pasadena, CA 91125, USA}
\affiliation{Center for Astrophysics, Harvard University, 60 Garden St., Cambridge, MA, 02138, USA}
\email{adolfo.carvalho@cfa.harvard.edu}

\begin{abstract}
We present near-infrared spectroscopic diagnostics that can be used to identify FU Orionis stars (FUOrs). 
FUOrs are young stellar objects (YSOs) that are currently in a state of extreme outburst, caused by enhanced mass {inflow} from their accretion disks. The disks give FUOrs a distinct multi-temperature optical and infrared spectrum. 
Considering both the predicted spectrum from a disk atmosphere model, and existing spectral diagnostics from the literature,
we identify key atomic and molecular features for characterizing FUOrs. 
Some of the chosen features are proxies for temperature, others are sensitive to surface gravity, and still others probe disk winds. 
Using the Palomar Observatory/Hale Telescope TripleSpec spectrograph, we gathered near-infrared spectra of 28 known FUOrs.  
We use standard equivalent widths to determine the strength of atomic lines and we design several band ratios for measuring molecular features. 
We compare the measurements between our spectra and a control sample of late-type dwarfs and evolved stars from the Infrared Telescope Facility Spectral Library. By considering the relative distributions of these samples in our defined spectral diagnostics, we propose a number of parameter spaces that can distinguish FUOr disks from normal stars. 
The rate of discovery of FUOr candidates has increased significantly in recent years,  
largely due to the increasing prevalence of time-domain surveys. 
Our proposed diagnostics will allow new photometric candidates to be confirmed or refuted as such.
\end{abstract}


\section{Introduction}\label{sec:intro}

Star formation begins with gravitational collapse of molecular clouds. As clouds become denser and hotter, they begin to radiate and form protostars surrounded by a {dusty} 
envelope. The protostar accretes as it contracts, gaining mass and becoming more luminous. It eventually forms a circumstellar disk and starts to blow off its envelope. The dissipation of dust allows the young stellar object to become optically visible. During this stage, it continues to gain mass until it appears on the HR diagram and transitions to a pre-main sequence star \citep{protostars}.


Stars can accrete as much as half of their main sequence mass from the envelope and disk during this early stage \citep{Le2024}. 
Unsurprisingly, variability is a very common trait among YSOs \citep{Fischer2023}, 
with FUOrs distinguished from other YSOs by the scale of this variability. 
FUOr outbursts derive from a massive, episodic {inflow }
of matter from their accretion disks. This gives a luminosity increase often on the scale of 3-4 magnitudes, and the sustained brightening can last for years to decades without much dimming \citep{Hartmann1996}.

While the large amplitude brightening is unique to FUOrs within the wider class of YSOs, there are other classes of variable stars that display similar brightening properties. For example, long period variables can have oscillations with rise times of years \citep{Suresh2024}, making partial lightcurves appear like FUOrs. 
Similarly, RCrB stars in the recovery phase from a deep fade can be confused with brightening FUOrs.
Other accretion-driven events such as symbiotic star outbursts can also be mistaken for FUOr outbursts based on lightcurves alone.
Therefore, spectroscopy must be used in conjunction with a photometric brightening in order to securely identify any new candidate FUOr.

The accretion outburst is radiating away gravitational energy, and the emission produced by the increased accretion rate through the inner disk contributes significantly to the spectrum,
creating a disk atmosphere that exhibits
an absorption line spectrum similar to that of a multi-temperature stellar atmosphere
\citep{Kenyon1987}. 
There is a radial temperature gradient with decreasing temperature moving outwards in the disk \citep{Kenyon1987, Welty1992} 
that can be effectively modeled as a collection of annuli where each has the spectrum of a stellar atmosphere corresponding to its radial distance from the star.  This $T(r)$ gradient gives FUOrs 
broad lines coming from the rotating accretion disk \citep{Petrov2008} with a multi-temperature spectrum that blends characteristics from both hotter and cooler stellar types \citep{Clarke2005}.

Their unique spectral fingerprint has long been used to identify FUOrs \citep{Hartmann1996, Connelly2018, Pena2019, Nagy2023, Hillenbrand2023}. However, many papers addressing this attribute of FUOrs have been more qualitative than quantitative. Spectra of new FUOr canidates are shown to appear similar to established FUOrs, sometimes pointing to one or two molecules as evidence but without any measurements. More rigorous identification leverages the mix of features referenced in \citet{Connelly2018}, which popularized equivalent width measurements of several nearby features (CO 2.30 $\mu$m,  \ion{Ca}{1} 2.26 $\mu$m, and \ion{Na}{1} 2.21 $\mu$m) to create plots that isolate FUOrs.  The diagnostic of CO 2.30 $\mu$m vs \ion{Na}{1} 2.21 $\mu$m $+$ \ion{Ca}{1} 2.26 $\mu$m   
was based on prior work by \citet{Greene1996} and \citet{Connelly2010}, which had the goal of identifying low-mass YSOs distinctly from both dwarfs and giants.  

We were thus motivated to conduct a more detailed investigation of the near-infrared spectra of FUOrs and produce further tests across the near-infrared spectral range capable of isolating FUOrs. 
In this paper, we identify and quantify the set of spectroscopic features that are typically present in FUOrs 
by studying a known population of widely accepted FUOrs. 
Our future goal is to apply our diagnostics to newly outbursting YSOs that are candidate FUOrs. 

In Section \ref{sec:acquisition}, we discuss new near-infrared observations of the sample of northern FUOr stars.
In Section \ref{sec:ref}, we describe a model disk spectrum that we use as reference for the ``ideal" FUOr spectrum (\S\ref{sec:model}) and also our use of the IRTF spectral library as empirical templates (\S\ref{sec:irtf}). 
In Section \ref{sec:dered}, we use the model to deredden the observed spectra so as to prepare them for the measurement of our spectral diagnostics. In Section \ref{sec:median} we define and describe the median FUOr spectrum. In Section \ref{sec:features}, we highlight the spectral features that we assess as meaningful diagnostics and explain why they were selected. In Section \ref{sec:ew}, we present the techniques used to quantitatively measure the strength of each identified feature. In Section \ref{sec:control}, we discuss the catalog of spectra used as non-FUOr sources. We then plot these stars against the FUOrs to distinguish between populations in Section \ref{sec:tests}.

\section{FUOr Sample and New Near-Infrared Spectroscopy} \label{sec:acquisition}

Our sample is comprised of sources classified as bona fide FUOrs or FUOr-like by \citet{Connelly2018}, as well as several more recently discovered FUOrs. {The former were selected by those authors from the literature, but assigned as bona fide FUOrs based on having an observed photometric outburst plus spectroscopic signatures including strong absorption from CO in K band, weak metal absorption, strong H$_2$O absorption in H-band creating a ``triangular" shape, and often having He I at 10830 \AA\ absorption.  FUOr-like sources are those meeting the spectroscopic criteria, but lacking evidence of an outburst.  The latter, more recently discovered FUOrs, come from the individual discovery papers of \citep{Siwak2025, Szegedi2020, Hillenbrand2018, Hillenbrand2021, Hillenbrand2023}.} The observed source list is shown in Appendix \ref{sec:datatable}, and a spectral atlas is provided in Appendix \ref{sec:objects}. 

Observations were taken on 19 and 20 January 2025 and on 11 and 12 July 2025 (UT). 
We employed the 200-inch Hale Telescope at Palomar Observatory and used 
the Triple Spectrograph \citep[TripleSpec;][]{Herter2008}, a cross-dispersed near-infrared spectrograph
covering the YJHK bands. 
TripleSpec has continuous wavelength coverage of 0.98-2.45 $\mu$m at spectral resolution $ \lambda/\Delta\lambda \equiv R \sim 2700$. Observations were taken using the standard ABBA dithering technique, with repeated sequences to build up signal-to-noise. {All observations were taken within 30 degrees of the parallactic angle and most were taken at airmass $<1.5$, so slit losses vs wavelength should not be significant.}

The spectra were flat-fielded, extracted, and co-added using using the \texttt{Spextool} package \citep{Cushing2004}. Telluric correction and flux-calibration were performed with \texttt{xtellcorr} \citep{Vacca2003} using A0V standard stars. The standards were observed several times over the course of each night so as to fully map airmass, with a telluric correction standard taken within 0.1 in sec$z$ of each science observation.
The final spectra have signal-to-noise exceeding 100 at wavelengths longer than 1.32 $\mu$m in about {60}\% of the objects.

\section{Reference Spectra}\label{sec:ref}

\begin{figure}[t]
\centering
\includegraphics[width=\linewidth]{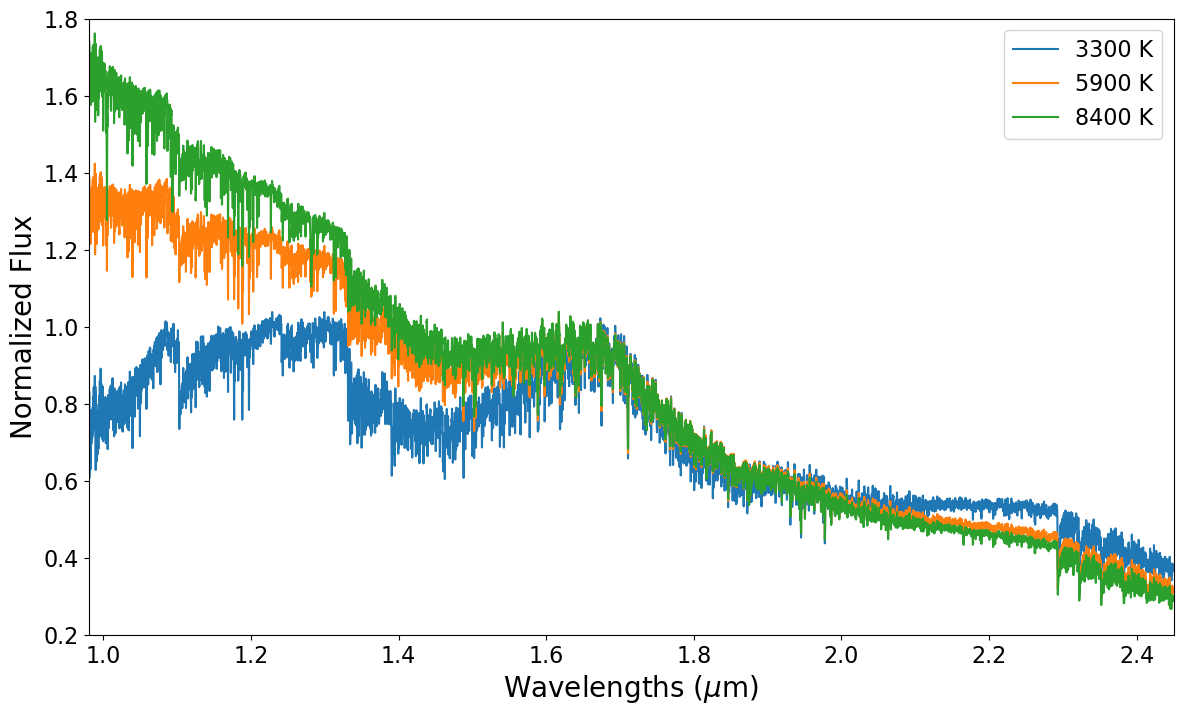}
\caption{Near infrared spectra for disk models having $T_{max}$ of 8400 K (green), 5900 K (orange), and 3300 K (blue), representing the hottest, intermediate, and coldest models considered here. 
The spectra are normalized at 1.67 $\mu$m. 
Both the overall spectral slope and the strength of various atomic and molecular features varies with temperature. 
We have adopted the 5900 K model in our dereddening procedure.}
\label{fig:all_models}
\end{figure}

\subsection{FU Ori Disk Model Atmosphere}\label{sec:model}


As a fiducial reference for the spectral energy distribution and detailed spectrum of FUOrs, 
we adopt the accretion disk model described
in \citet{Carvalho2024}, which was fit to FU Ori itself. The model assumes a geometrically thin disk that is viscously heated, with radiative dissipation of that heat. The radial temperature profile of the disk is given by
\begin{equation}
    T_\mathrm{eff}(r) = \left[ \frac{3 G M_* \dot{M}}{8 \pi \sigma_{SB} r^3} \left(1 - \sqrt{\frac{R_\mathrm{inner}}{r}} \right) \right]^{1/4},
\end{equation}
for $r \geq \frac{49}{36} R_\mathrm{inner}$ and $T_\mathrm{eff}(r < \frac{49}{36} R_\mathrm{inner}) = T_\mathrm{eff}(\frac{49}{36} R_\mathrm{inner}) \equiv T_\mathrm{max}$. $M_*$ is the mass of the central star, $R_\mathrm{inner}$ is the inner boundary of the disk, $\dot{M}$ is the disk-to-star accretion rate, $G$ is the universal gravitational constant, and $\sigma_{SB}$ is the Stefan-Boltzmann constant. We use the best-fit parameters of FU Ori itself from \citet{Carvalho2024}: $M_* = 0.6 \ M_\odot$, $\dot{M} = 10^{-4.5}$, and $R_\mathrm{inner} = 3.52 \ R_\odot$, which yield $T_\mathrm{max} \approx 5900$ K. 

The model spectrum is computed by discretizing the disk into annuli and assigning a BT-Settl stellar atmosphere model \citep{Allard2013} of the appropriate $T_\mathrm{eff}$ to each annulus. The spectra are then combined via an area-weighted sum to produce a final disk spectrum. To match the resolution of the observations, the spectra are convolved with a Gaussian kernel whose standard deviation is equal to the quadrature difference between the native resolution of the models and the observations.

We also computed a grid of accretion disk models spanning a wide range of $T_\mathrm{max}$ from 3300 K to 8400 K. Selected spectra from this model grid are shown in Figure \ref{fig:all_models}. 
The set of model spectra show a relatively small color range, 
and there are also differences in the strengths of the spectral features. 
Among individual FUOr sources, $T_{max}$ values do vary, 
but over a much narrower range than our model grid, and
a single model was chosen for use in this study.
We adopt $T_{max} \approx 5900$ K, 
based on the distribution of $T_{max}$ values found in  \cite{CarvalhoThesis2025}. 
Alternative $T_{max}$ values over the range defined in Figure~\ref{fig:all_models} would have a systematic effect on the $A_V$ that we derive. However, the changes in $A_V$
remain within our reported errors, and have a negligible impact on the reported spectral line
and band strengths.

\subsection{Empirical Standards}\label{sec:irtf}

We also make use of the NASA Infrared Telescope Facility (IRTF) Spectral Library\footnote{\url{https://irtfweb.ifa.hawaii.edu/~spex/IRTF_Spectral_Library/}} \citep{Rayner2009}. 
This catalog contains near-infrared spectra for stars ranging over luminosity classes I-VI and over spectral types F-M.  Dwarf types L and T are also included, as are evolved luminous S/C-type stars.
 
In practice, we restrict our comparisons to the M types and the S stars,
since these are the spectra that could be most easily confused with the composite FUOr disk atmosphere spectra. This decision is discussed in \S\ref{sec:control}.

\section{Analysis}\label{sec:methods}
\subsection{FUOr Dereddening Procedure}\label{sec:dered}

\begin{figure}[b]
\centering
\includegraphics[width=1.1\linewidth]{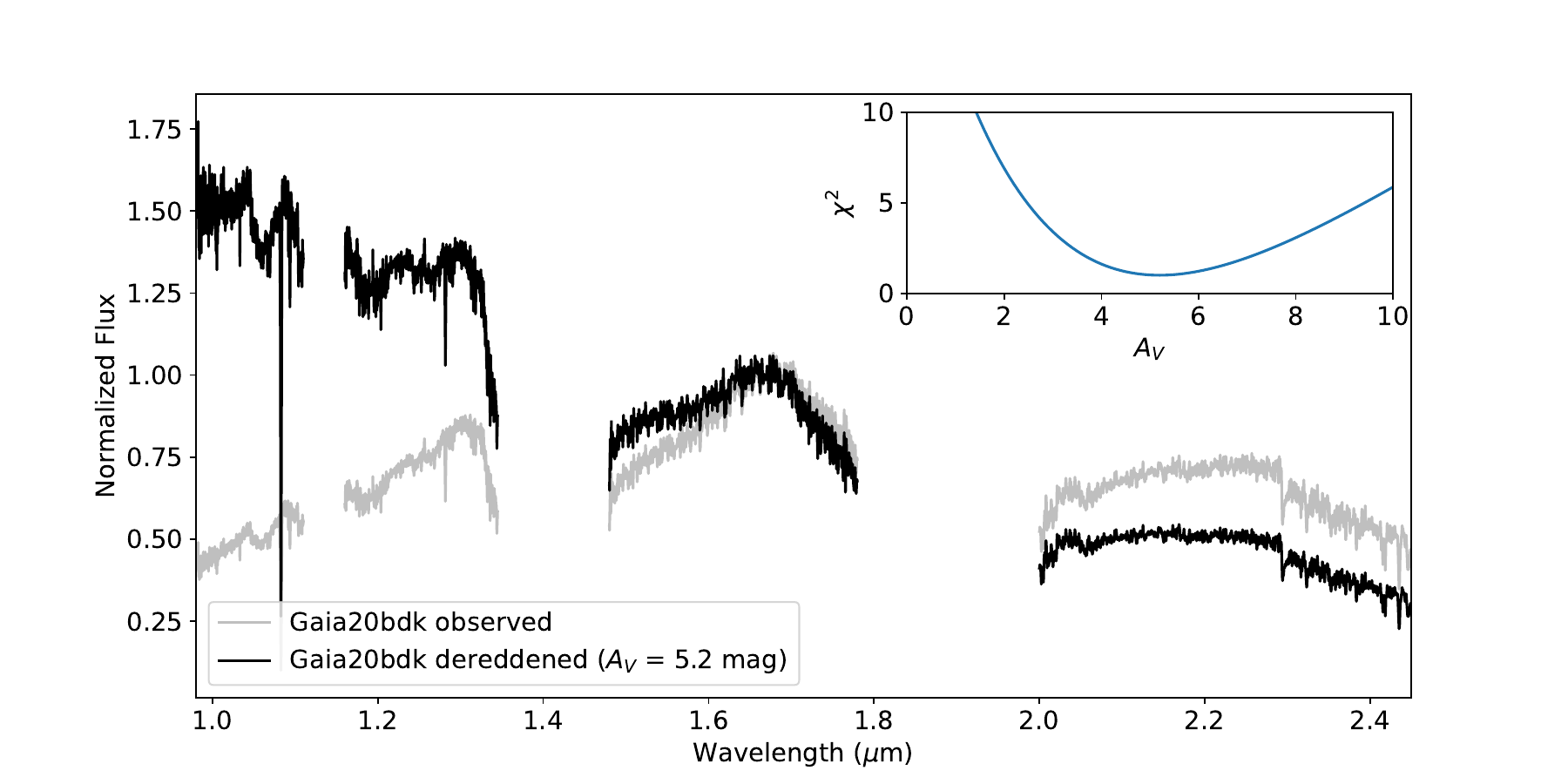}
\caption{
Example of the observed (gray) and resulting dereddened (black) spectra, in the case of Gaia 20bdk.
Both spectra are normalized at 1.67 $\mu$m. 
Inset shows the $\chi^2$ curve that produces the $A_V$ for the source, normalized at the minimum.
{\it Figure is part of a Figure set available for each of the 28 sources in our sample.}
}
\label{fig:atlas}
\end{figure}

The observed spectra of FUOr sources are usually reddened, since these sources are typically found
in heavily dust-extincted regions within molecular clouds.  
Substantial reddening can affect measurements of spectral features,
especially molecular bands. We thus implement a dereddening procedure for our spectral data,
with results shown in Figure~\ref{fig:atlas}.
Reddening also differentially reduces spectral SNR, with the largest impact towards bluer wavelengths.
We account for this below by {incorporating a weighting by} error in our dereddening fits.

\begin{figure}[t]
\centering
\includegraphics[width=\linewidth]{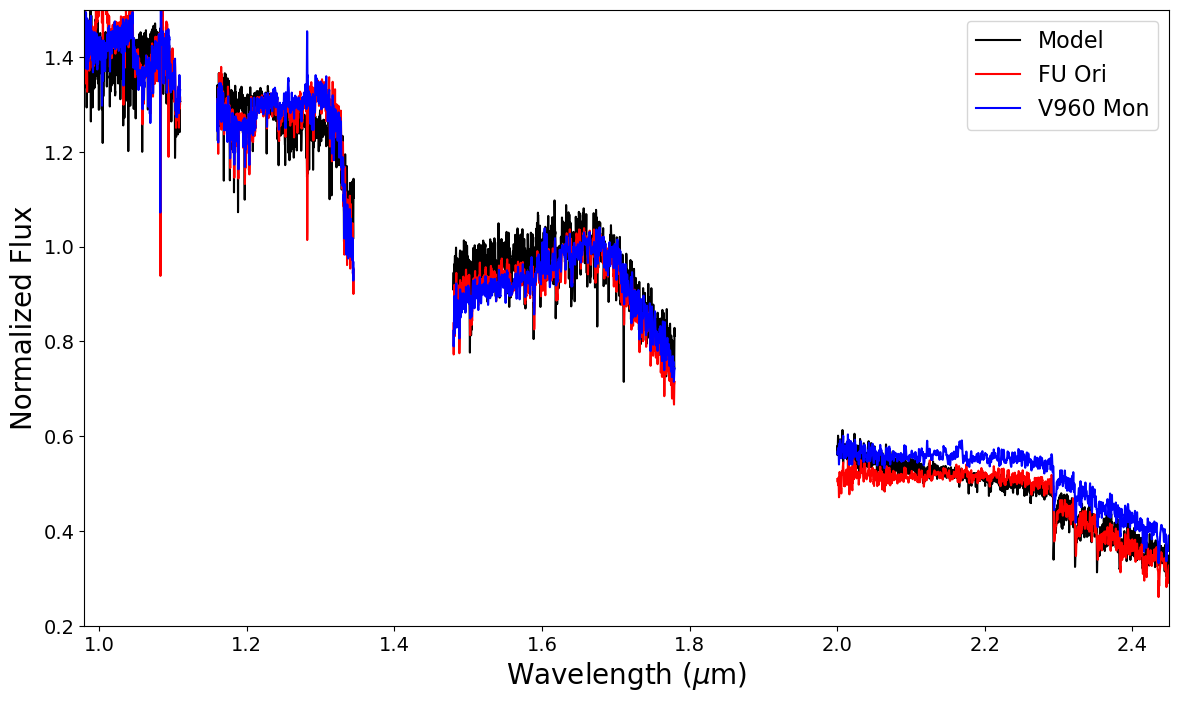}
\caption{
Comparison of dereddening results for two sources, FU Ori (red) and V960 Mon (blue),
shown on top of the disk model (black) to which the observed spectra were dereddened. 
All spectra are normalized at 1.67 $\mu$m. 
The overall match is excellent over the YJH region, albeit with some variation. 
At K band, while FU Ori is quite well-modeled, the V960 Mon flux is significantly brighter than in the model. 
Even though a K-band excess over the disk model is not always present,
we systematically disregard the K-band region in our dereddening procedure, fitting only the YJH region.
}
\label{fig:dereddened}
\end{figure}



Our observations are most useful within the following wavelength ranges: 
$0.98-1.11~\mu$m (Y band), $1.16-1.345~\mu$m (J band), $1.48-1.78~\mu$m (H band), and $2.0-2.45~\mu$m (K band). 
Areas between these bands are heavily impacted by  poor atmospheric transmission and therefore have particularly low SNR. 
The inter-band regions are universally unusable and are omitted from our analysis. 

Before dereddening, 
we normalize both the $T_{max} \approx 5900$ K disk model and the observed spectra to the middle of the H band at 1.67 $\mu$m. 
We choose H band because the bluer Y and J bands can have significant noise due to source redness. On the red side, K band can have flux contributions above the accretion disk model from dust emission, as discussed below.  
The wavelength of 1.67 $\mu$m was selected because it is the peak of the ``elbow" feature, so it will be least affected by the H$_2$O that depresses flux on either side of the peak. To mitigate the effects of noise in H band,
we use the median value over 1.66-1.68 $\mu$m to determine the normalization value.

Our dereddening procedure uses only Y, J, and H bands. We do not include K band in determining $A_V$.
Observationally, the flux at K band relative to that at H, J, and Y bands 
varies widely within the population of known FUOrs.
This is a result of source-to-source differences in the contribution from thermal emission 
arising in the outer passive disk as opposed to the inner active disk \citep{Liu2022}. 
{Specifically, the outer parts of the disk are passively heated 
due to dust irradiation, while the innermost disk is actively heated due to viscous accretion.  
This dust component is not included in our disk model, which is of the gaseous accretion disk only.
The Wien side of the dust component can become an important contributor to the total flux 
of some FUOrs somewhere between about 2 and 5 $\mu$m.}   
Although the effect is stronger in the mid-infrared, it can be present as blue as K band.
The inconsistency among FUOrs in the presence and strength of the ``K excess" can bias the best-fit $A_V$.
{Following the procedure described below, we find 30\% of the sources in our catalog likely contain this excess.}
{Furthermore, as source redness can cause excessive noise at the shorter wavelengths, 
when the signal-to-noise in J band is below 20, we ignore spectral regions shortward of 1.48 $\mu$m.} 

With the above pre-processing steps completed, we apply the dereddening code from the \texttt{dust\_extinction} library \citep{Gordon2024}. We adopt the \citet{Gordon2009} reddening law derived from measurements of the galactic disk spanning near-infrared wavelengths, as implemented in \texttt{GCC09\_MWAvg}.
Procedurally,
we iterate through $A_V$ values from 0-40 mag in steps of 0.1 mag, applying each to the observed spectrum before comparing it to the model. The goodness-of-fit is tested through a reduced $\chi^2$ statistic 
on an interpolated wavelength grid using 
\begin{equation}\label{eq:chi}
    \chi^2=\frac{1}{N-1}\sum_\lambda\left(\frac{F_{\lambda ,o}-F_{\lambda, m}}{\sigma_\lambda}\right)^2.
\end{equation}
We report the $A_V$ that returns the minimum $\chi^2$. The error bars are  
estimated by finding the $A_V$ range over which the $\chi^2$ value increases to twice its minimum value, {after adopting min$(\chi^2)=1$; \citealt{Andrae2010}}. 
A comparison of raw and dereddened spectra along with the $A_V$ fit is shown in Figure \ref{fig:atlas}.

Figure \ref{fig:dereddened} illustrates the dereddened spectra for two cases with differing amounts of thermal emission contributing to the K-band flux, rather than only extinction affecting the spectrum.
Were we to consider K band while dereddening, there would be a seesaw effect about our normalization wavelength of 1.67 $\mu$m, in which the dereddened K band would decrease while Y/J bands would increase for higher values of $A_V$, and vice versa for lower values of $A_V$. This would result in a dereddened spectrum where both ends of the spectrum were poorly fit and an overall increase in $\chi^2$.

To validate our methods, we tested them on the data of \citet{Connelly2018}.  
Beyond removing regions between the atmospheric bands with poor transmission, we mirrored the SNR cuts applied by these authors, for consistency. Out of 31 spectra, this caused 9 to lose their Y-band data, 5 of which also lost part of J band. As we also disregard K band while dereddening, there is little spectrum to go on for some sources, and thus our A$_V$ estimates may have less reliability for this sub-sample of the \citet{Connelly2018} data set.

\begin{figure}
\centering
\includegraphics[width=0.49\textwidth]{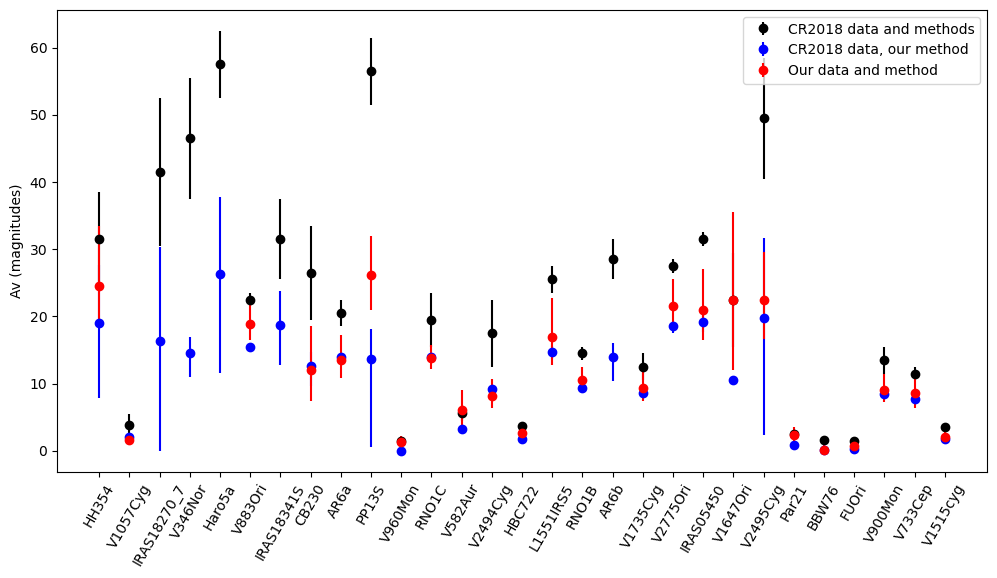}
\caption{\label{fig:av_comp}
Comparison of $A_V$ values derived in this paper with those from \citet{Connelly2018}. Black points are reported in \citet{Connelly2018}. Blue points are found by using our dereddening method on the \citet{Connelly2018} dataset. Red points show the $A_V$ values reported in Appendix \ref{sec:datatable} using the new observations from \S\ref{sec:acquisition} and the dereddening methods described
in \S\ref{sec:dered}. Our $A_V$ values are generally lower than those of \citet{Connelly2018}. 
}
\end{figure}

The results of our dereddening comparison are shown in Figure \ref{fig:av_comp}. In general, our methods produce the same overarching trends as \citet{Connelly2018}, in terms of more heavily reddened vs lightly reddened sources. However, our $A_V$ values are consistently lower due to differences in methodology, with
several potential contributing factors. 

Regarding the spectral fitting, first, the \citet{Connelly2018} dereddening fit uses the entire YJHK spectrum 
and does not consider the possibility of a K excess. This results in systematically higher $A_V$ than we infer. 
Second, \citet{Connelly2018} normalize in K band. 
This would make the bluer parts of the spectrum appear fainter relative to our H band normalization, and thus require more dereddening to match observations. Third, \citet{Connelly2018}
used an observed FU Ori spectrum as the baseline for dereddening, rather than a disk atmosphere model,
introducing additional noise in the dereddening process.


The methodology differences also extend to the application and evaluation of the extinction. \citet{Connelly2018} adopted the method of \citet{Connelly2010} using a power law extinction model based on observations toward the Galactic center. The model we adopt from \texttt{dust\_extinction} draws from observations along multiple Galactic sightlines, and is functionally more complex, but in practice is similar in the near-infrared wavelength range. 
The appropriateness of any given extinction law for YSOs depends on their location,
and whether in the Galactic plane and distant, where the extinction is dominated by the diffuse ISM, or in an out-of-plane nearby molecular cloud, where the extinction is dominated by the dense ISM.
In working with the \citet{Connelly2018} data, for which we do not have measured errors, 
we use the following reduced $\chi^2$ equation instead of Equation \ref{eq:chi}, 
\begin{equation}
    \chi^2=\frac{1}{N-1}\sum_\lambda\frac{(F_{\lambda ,o}-F_{\lambda, m})^2}{F_{\lambda ,m}}.
\end{equation}

\begin{figure*}
\centering
\includegraphics[width=\textwidth]{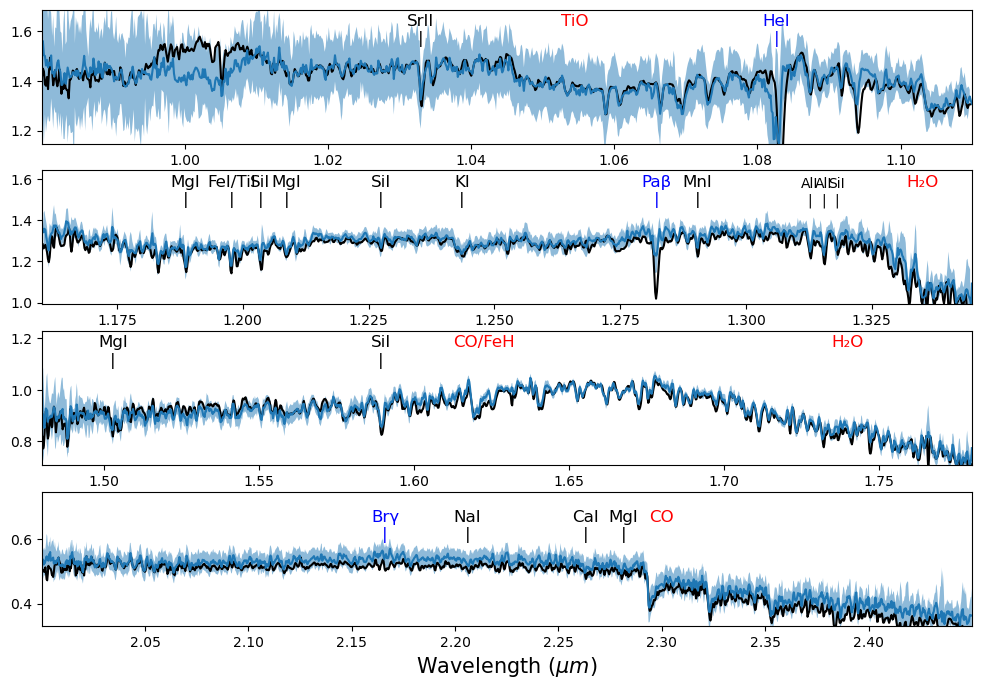}
\caption{\label{fig:median}
The sigma-clipped median flux (blue line) and median absolute deviation (blue shaded region)
constructed after applying our dereddening procedure to all observed FUOrs in our sample. 
This is  compared to the spectrum of the prototype FU Ori itself (black). 
Spectral features that we measure are labeled, with atomic lines in black, molecules in red, and wind-sensitive lines in blue. 
The excellent match of the black and blue lines shows that FU Ori itself accurately represents the class.
The low dispersion also shows that most of the spectral features are shared by all FUOrs. 
Increased scatter towards Y band is largely due to decreasing signal-to-noise at the blue end of our spectral coverage and is considered insignificant.  
Increased scatter towards K band is likely real and due to differing contributions among the sources from the passive (dust) disk. 
Scatter can also increase away from the wavelength of normalization (1.67 $\mu$m) due to errors in the $A_V$ correction.    
{\it The median FUOr spectrum is available as a data-behind-the-figure file.}
}
\end{figure*}

\subsection{FUOr Median YJHK Spectrum}\label{sec:median}
To assess our entire spectroscopic catalog, we first create a "median" FUOr spectrum. 
After normalizing and dereddening each individual spectrum using the techniques from \S\ref{sec:dered}, we combine the spectra via the sigma-clipped median in each wavelength bin. This minimizes the impact of spectra with very low signal-to-noise in the bluer bands. We also compute the median absolute deviation in each wavelength bin, which we treat as the characteristic spread among the sample.
This median spectrum is displayed in Figure \ref{fig:median}, alongside FU Ori itself.   
FU Ori and the median spectrum are almost identical, differing mainly in the wind-dominated lines of
\ion{He}{1} 1.08 $\mu$m, Pa$\beta$ 1.28 $\mu$m, plus other higher hydrogen lines that are unmarked.
The excellent match justifies our use of the SED provided by the FU Ori model in our dereddening procedure.

The median FUOr spectrum in Figure \ref{fig:median} is shaded according to the dispersion among the observed FUOrs. There are two areas with higher than typical dispersion. The first is in K band around the CO 2.30 $\mu$m lines. This is further evidence of the excess continuum contributing variably among the FUOr population, as discussed in \S\ref{sec:dered}. The other region with high dispersion is Y band, likely due to the extreme redness of FUOrs and consequent low SNR, which is less than 10 in some cases.  

\subsection{FUOr Spectral Feature Selection}\label{sec:features}

FUOr spectra have a mix of strong molecular and atomic absorption features, many of which are effective probes for parameters such as temperature, surface gravity, or wind/outflow. In typical stellar spectra, the temperature-sensitive absorption lines indicate a single effective temperature for the star (or possibly two if there are large, cool spots on the stellar surface). However, in FUOrs, the spectrum is dominated by the disk atmosphere with decreasing temperature as a function of radius. The resulting observed spectrum is a mix of spectral types, and different wavelength ranges yield inconsistent effective temperature measurements \citep{Herbig1977}. This distinction between FUOr disks and more typically accreting YSOs motivates the creation of diagnostics that uniquely identify FUOrs spectroscopically. 

\begin{deluxetable}{c|cc|cc|cc}
\tablewidth{\linewidth}
\tablecaption{Measured Features \label{tab:all_features}}
\tablehead{
\multicolumn{1}{c}{Feature} &  \multicolumn{2}{|c}{Continuum 1} & \multicolumn{2}{|c}{Measurement Range} & \multicolumn{2}{|c}{Continuum 2}\\
&          $\lambda_{start}$ &          $\lambda_{end}$ & 
         $\lambda_{start}$ &          $\lambda_{end}$ & 
         $\lambda_{start}$ &          $\lambda_{end}$ \\
& [$\mu m$] &  [$\mu m$] & [$\mu m$] &  [$\mu m$] & [$\mu m$] &  [$\mu m$]}
\startdata
\multicolumn{7}{c}{Y Band}\\ 
\hline
\ion{Sr}{2} & 1.0309 & 1.0319 & 1.0321 & 1.0340  & 1.0359 & 1.0369 \\
TiO & 1.0405 & 1.0420 & 1.0545 & 1.0575 &\nodata &\nodata\\
\ion{He}{1} & 1.0762 & 1.0775 & 1.0805 & 1.0848 & 1.0848 & 1.0862 \\
\hline
\multicolumn{7}{c}{J Band}\\ 
\hline
\ion{Mg}{1} & 1.1860 & 1.1880 & 1.1880 & 1.1893 & 1.1915 & 1.1925 \\
\ion{Fe}{1}/\ion{Ti}{1} & 1.1962 & 1.1969  & 1.1969  & 1.1984  & 1.2012 & 1.2022 \\
\ion{Si}{1} & 1.2012 & 1.2022 & 1.2026 & 1.2046 & 1.2064 & 1.2074 \\
\ion{Mg}{1} & 1.2064 & 1.2074 & 1.2077 & 1.2095 & 1.2144 & 1.2154 \\
\ion{Si}{1} & 1.2200 & 1.2217 & 1.227 & 1.2283 & 1.231 & 1.233 \\
\ion{K}{1} & 1.2370 & 1.2385 & 1.2415 & 1.2453 & 1.2464 & 1.2477 \\
Pa $\beta$ & 1.2785 & 1.2795 & 1.2813 & 1.283 & 1.2862 & 1.2872 \\
\ion{Mn}{1} & 1.2862 & 1.287 & 1.2891 & 1.2916 & 1.2918 & 1.2926 \\
\ion{Al}{1} & 1.3044 & 1.3054 & 1.3113 & 1.3138 & 1.3194 & 1.3204 \\
\ion{Al}{1} & 1.3044 & 1.3054 & 1.3144 & 1.3166 & 1.3194 & 1.3204 \\
H$_2$O & 1.2950 & 1.3050 & 1.3350 & 1.3450 &\nodata&\nodata\\
\hline
\multicolumn{7}{c}{H Band}\\ 
\hline
\ion{Mg}{1} & 1.4970 & 1.5 & 1.5010 & 1.5060 & 1.5060 & 1.5090 \\
\ion{Si}{1} & 1.5852 & 1.5867 & 1.5870 & 1.5910 & 1.5930  & 1.5945 \\
CO/FeH & 1.6162 & 1.6177 & 1.6175 & 1.6280 & 1.6275 & 1.6290 \\
H$_2$O & 1.6600  & 1.6800 & 1.7200 & 1.7600 &\nodata&\nodata\\
\hline
\multicolumn{7}{c}{K Band}\\ 
\hline
Br $\gamma$ & 2.1635 & 2.1646 & 2.1646 & 2.1673 & 2.1673 & 2.1678 \\
\ion{Na}{1} & 2.1825 & 2.1890 & 2.2040 & 2.2110 & 2.2140 & 2.220 \\
\ion{Ca}{1} & 2.2490 & 2.2530 & 2.2580 & 2.2690 & 2.2700 & 2.2740 \\
\ion{Mg}{1} & 2.2720 & 2.2760 & 2.2760 & 2.2850 & 2.2850 & 2.2890 \\
CO & 2.2825 & 2.2900 & 2.2850 & 2.3165 & \nodata & \nodata \\
\enddata
\tablecomments{The values used to define each measured feature from \S\ref{sec:features}. 
}
\end{deluxetable}

The features that we select must be shared by the entire population for them to be useful diagnostics.
We thus searched the median FUOr spectrum to select viable features for measurement. Measurable features are strong compared to typical continuum noise and are relatively isolated, allowing adjacent continuum to be accurately defined. 
The features that we consider measurable are marked in Figure \ref{fig:median} and defined in
Table \ref{tab:all_features}.


In K band, \citet{Connelly2018} showed that CO 2.30 $\mu$m is the strongest molecular feature and that it, \ion{Na}{1} 2.21 $\mu$m, and \ion{Ca}{1} 2.26 $\mu$m are good spectral diagnostics in K band. We take the wavelengths encompassing the features and defining the continuum region from \citet{Messineo2021}, with some adjustment. \citet{Messineo2021} also include the relatively strong \ion{Mg}{1} 2.28 $\mu$m feature, which we also measure. Br$\gamma$ 2.17 $\mu$m is selected as a wind line. 

In H band,
we measure the strong \ion{Si}{1} 1.59 $\mu$m line using continuum regions as defined by \citet{Messineo2021}, and we also measure the \ion{Mg}{1} 1.50 $\mu$m doublet. 
A strong molecular feature near the middle of H band is FeH 1.62 $\mu$m\footnote{
The feature that we have measured and are calling FeH is actually a combination of three molecular bands: a CO $\delta\nu=3$ bandhead at 1.6189 $\mu$m, several OH $\delta\nu=2$ bands around 1.62 $\mu$m, and FeH at 1.6250  $\mu$m; see materials in \citet{Cushing2005,Rayner2009}.},
for which we have again adopted the continuum regions defined by \citet{Messineo2021}.
We also measure H$_2$O 1.74 $\mu$m that depresses the red side of the H band, redward of the peak where we have normalized the spectra (\S \ref{sec:dered}).

In J band, we adopt the H$_2$O index at 1.34 $\mu$m as defined by \cite{Slesnick2004}, 
and we also measured several atomic lines. 
In Y band, we selected TiO 1.05 $\mu$m (though not typically very strong, it is a broad and distinguishing feature) as a molecular and also the \ion{Sr}{2} ion.
We also measured the wind-diagnostic lines \ion{He}{1} 1.08 $\mu$m in Y band and Pa$\beta$ 1.28 $\mu$m in J band. 


\subsection{Feature Strength Measurements}\label{sec:ew}

We used three different methods to measure feature strengths. The vast majority are equivalent width (EW) measurements. One feature employs a variant on EW integration, and three other features use a spectral index.

Our EW measurements are made using the standard equation 
\begin{equation}
    W_\lambda=\int^{\lambda_2}_{\lambda_1}\frac{F_c(\lambda)-F(\lambda)}{F_c(\lambda)}d\lambda.
\end{equation}

The error calculations for the EWs are taken from  \citet{Vollman2006} and based on error propagation via Taylor expansion:
\begin{equation}
    \sigma(W_\lambda)=\sqrt{\left(\frac{\Delta\lambda}{\overline{F_c}}*\sigma(F)\right)^2+\left(\frac{\Delta\lambda-W_\lambda}{\overline{F_c}}*\sigma(F_c)\right)^2}.
\end{equation}

The choice of boundaries for both the bands and the surrounding continua were made largely through empirical consideration with some input from the literature (as discussed in \S\ref{sec:features}) and are provided in Table \ref{tab:all_features}.

For standard ``two-sided EW" measurements, we select boundaries for the continuum on either side of the feature. We assume, in the absence of any noise, that the flux of the continuum would be a straight line given the small wavelength range. This line connects the average flux of the defined continua surrounding the feature. This allows line strength to be quantified by comparing the feature flux to the extrapolated continuum.

The CO 2.30 $\mu$m bandhead is a unique spectral feature, in that there are multiple successive broad absorption regions with no observed continuum on the red side. 
For this line, we adopt a ``one-sided EW" technique by taking continuum on the blue side of the feature only and extrapolating this continuum over the feature for the integration, following \citet{Messineo2021}.

TiO 1.05 $\mu$m, H$_2$O 1.34 $\mu$m, and H$_2$O 1.74 $\mu$m also require special treatment.  For these features we adopt a spectral index rather than an EW technique.  These molecules absorb over a broad range of wavelengths rather than being narrow Gaussian-like absorptions, and we cannot effectively integrate across the full features. Our spectral indices consist of choosing continuum on the blue side and taking a ratio of that continuum's average with the depressed flux of the feature, as in \citet{Slesnick2004}.  A spectral index thus has a value of unity for no signature of the feature and values $<1$ for stronger features. 

\begin{figure}
\centering
\includegraphics[width=\linewidth]{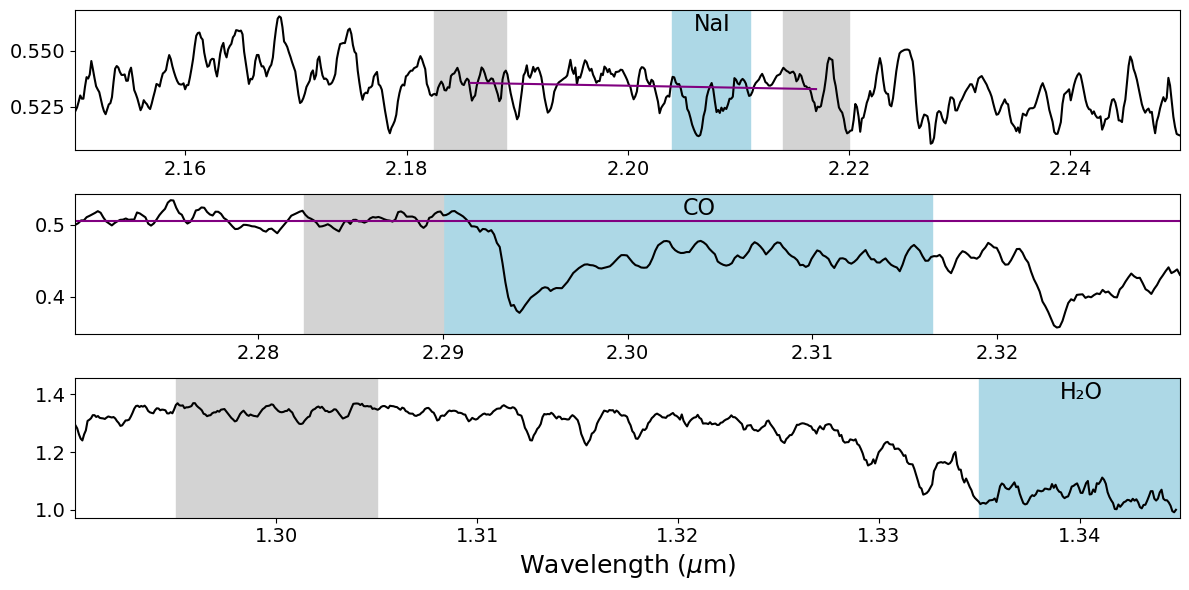}
\caption{\label{fig:ew_methods}
Illustration of each of our three spectral feature measurement techniques conducted on the median FUOr spectrum from Figure \ref{fig:median}. Blue shaded regions encompass the targeted feature, while gray shaded regions show the continuum region(s). The purple line represents the fitted continuum as described in \S\ref{sec:ew}. The top panel shows the \ion{Na}{1} 2.21 $\mu$m doublet as an example of a standard two-sided EW technique. The middle panel shows CO 2.30 $\mu$m, a one-sided EW technique. The bottom panel shows H$_2$O 1.34 $\mu$m as an example of a spectral index. A full catalog showing the measurement ranges of each feature can be found in Appendix \ref{sec:shading}.}
\end{figure}

Examples of each of our three types of measurements are shown in Figure \ref{fig:ew_methods}. 
The full set of  spectral line or band features and their respective continuum regions are shown for every measured feature in Appendix \ref{sec:shading}.
The measured values of these features for all FUOrs in our sample can be found in Appendix \ref{sec:datatable}. 
{The full datatable for the IRTF catalog is given as a \textit{data-behind-the-figure} file.}

It is possible that we are occasionally underestimating some EWs in this process. In some lower SNR sources, the spectrum is too contaminated with noise to find a realistic continuum. Similarly, when we look at some more evolved stars in \S\ref{sec:control}, our defined measurements may be somewhat compromised since these sources have more absorption around our defined continuum regions than the typical dwarfs, giants, and supergiants. 

\begin{figure*}[ht!]
\centering
\includegraphics[width=0.9\textwidth]{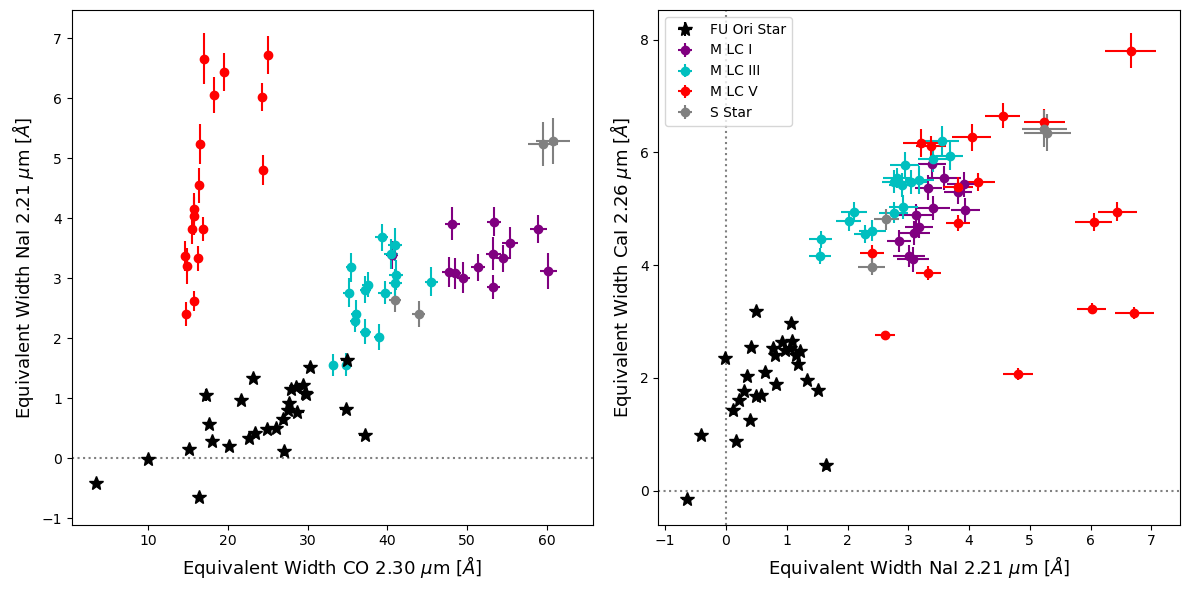}
\caption{\label{fig:k_band}
FUOr diagnostics in K band. On the left, a plot of CO 2.30 $\mu$m vs \ion{Na}{1} 2.21 $\mu$m EWs compares a strong molecule with an atomic feature. On the right, a plot of \ion{Na}{1} 2.21 $\mu$m vs \ion{Ca}{1} 2.26 $\mu$m EWs compares two atomic features. FUOrs are weaker in both atoms compared to standard M-type photospheres at any luminosity, including S-type sources. \ion{Na}{1} 2.21 $\mu$m in particular increases towards lower temperatures and higher surface gravities, explaining why 
lower gravity FUOrs are weak in \ion{Na}{1} 2.21 $\mu$m. CO 2.30 $\mu$m strength grows weakly with decreasing temperature but is also strongly inversely proportional to surface gravity, explaining why FUOrs have intermediate CO 2.30 $\mu$m strength, generally between dwarfs and giants.}
\end{figure*}

\begin{figure*}[hb!]
\centering
\includegraphics[width=0.9\textwidth]{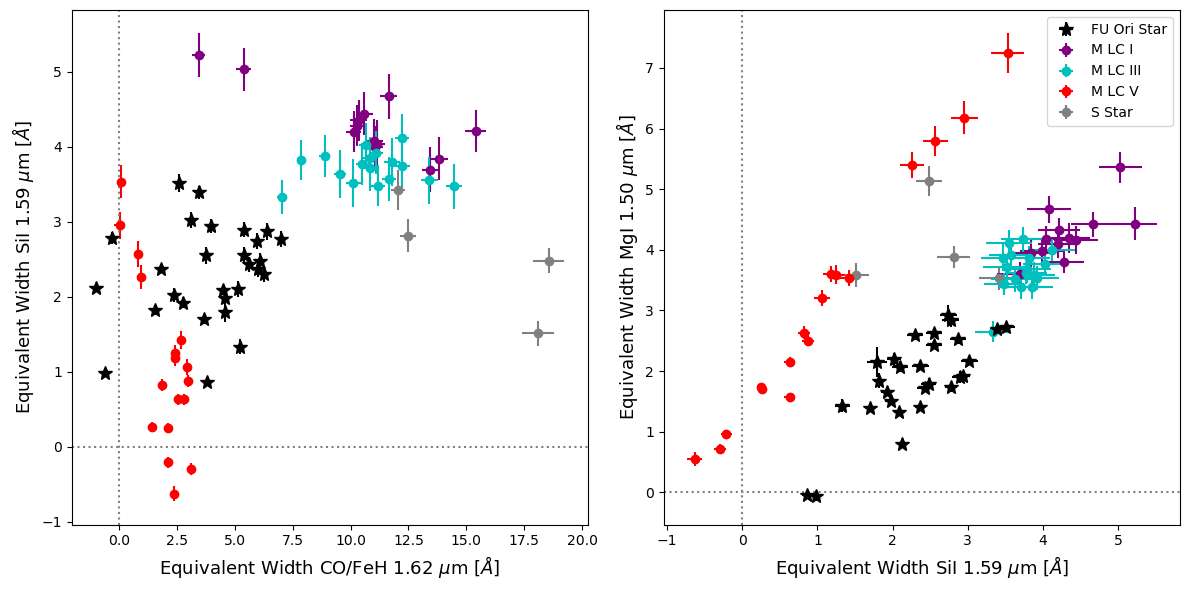}
\caption{\label{fig:h_band}
FUOr diagnostics in H band. 
On the left, a plot of FeH 1.62 $\mu$m vs \ion{Si}{1} 1.59 $\mu$m EWs compares a strong molecule with an atomic feature. 
The FeH 1.62 $\mu$m strength has a weak temperature dependence, and the gap between giants and dwarfs is due to the feature strength decreasing with increasing surface gravity.
FUOrs have intermediate FeH 1.62 $\mu$m and also \ion{Si}{1} 1.59 $\mu$m 
that is on average stronger than dwarfs and weaker than giants.
On the right, a plot of \ion{Si}{1} 1.59 $\mu$m vs \ion{Mg}{1} 1.50 $\mu$m EWs compares two atomic features. 
For the M dwarfs, this parameter space shows a linear sequence with the feature strengths peaking at late K/early M spectral types and decreasing towards cooler M types; the S-type sources also reside along this sequence. The FUOrs have similar \ion{Mg}{1} 1.50 $\mu$m line strength as the M dwarfs, but
as noted above, stronger \ion{Si}{1} 1.59 $\mu$m compared to the dwarfs, residing closer to but weaker than the giants.
}
\end{figure*}

\begin{figure*}[ht!]
\centering
\includegraphics[width=0.9\textwidth]{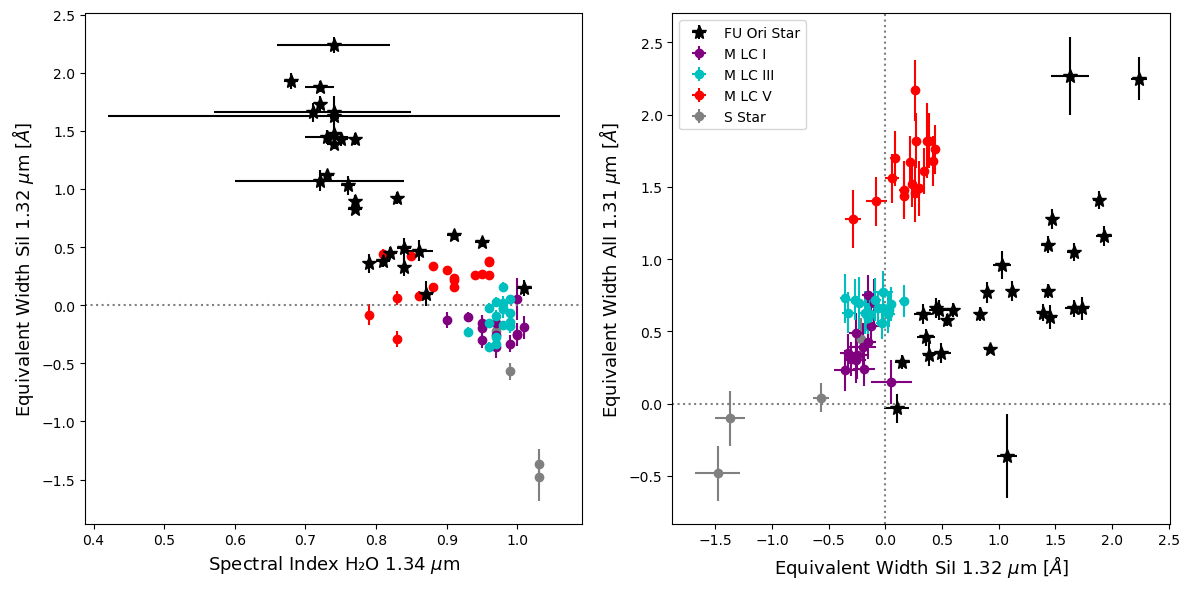}
\caption{\label{fig:j_band}
FUOr diagnostics in J band. On the left, a plot of H$_2$O 1.34 $\mu$m spectral index vs \ion{Si}{1} 1.32 $\mu$m EW, comparing a strong molecule with an atomic feature. The H$_2$O 1.34 $\mu$m signature (unity is no detectable H$_2$O) is more prominent at lower temperatures and is weakly dependent on surface gravity, being stronger in dwarfs than in lower gravity giants.
FUOrs are strong in both H$_2$O 1.34 $\mu$m and SiI 1.32 $\mu$m absorption, and stand apart from M-type stars at all luminosities, as well as S-type sources (which may be poorly measured).
On the right, a plot of \ion{Si}{1} 1.32 $\mu$m vs \ion{Al}{1} 1.31 $\mu$m EWs, comparing two atomic features
that are right next to one another in the spectra.  The \ion{Al}{1} 1.31 $\mu$m feature has a strong surface gravity dependence (being stronger in dwarfs than in giants) and grows in strength towards the coolest M-type temperatures.  The \ion{Si}{1} 1.32 $\mu$m feature is also stronger in dwarfs than in giants.
Similar to the case of \ion{Si}{1} 1.59 $\mu$m,  FUOrs have generally stronger \ion{Si}{1} 1.32 $\mu$m 
than either dwarfs or giants and weaker \ion{Al}{1} 1.31 $\mu$m than M dwarfs, though they somewhat overlap with the M giants.
}
\end{figure*}

\begin{figure*}[hb!]
\centering
\includegraphics[width=0.98\textwidth]{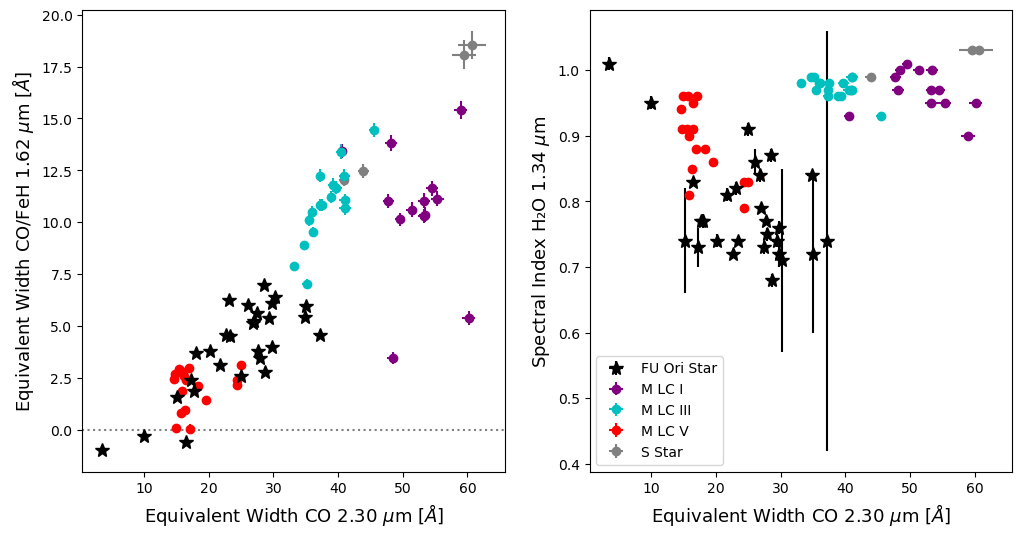}
\\
\includegraphics[width=0.49\textwidth]{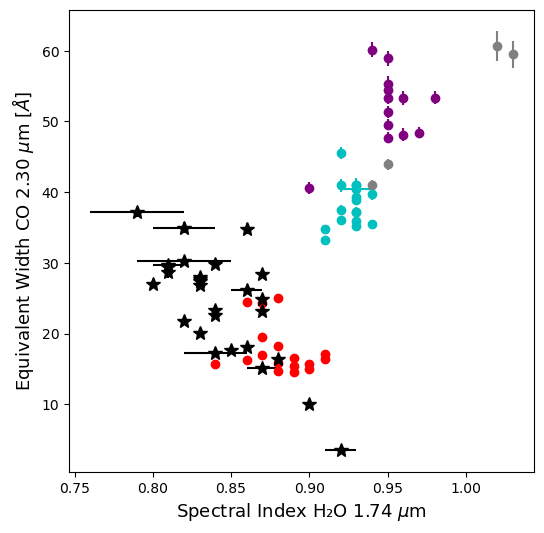}
\includegraphics[width=0.49\textwidth]{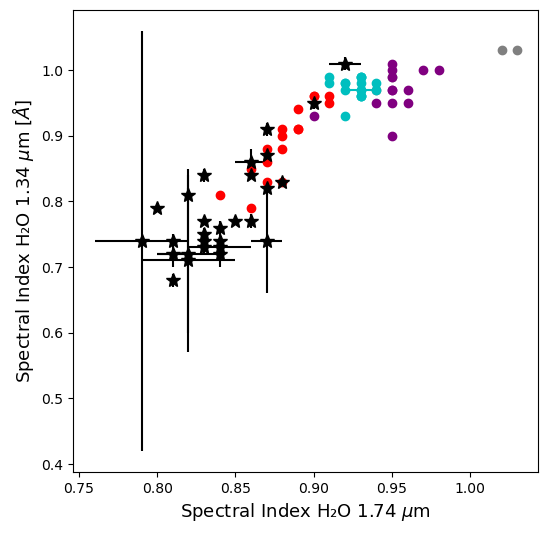}
\caption{\label{fig:crossmolecule}
Comparison of FUOr molecular diagnostics. 
In the top panels, the CO 2.30 $\mu$m EW in K band is plotted vs FeH 1.62 $\mu$m EW in H band (left) and vs the H$_2$O 1.34 $\mu$m spectral index in J band (right). Both plots show some overlap between FUOrs and M dwarfs, with the FUOrs tending to be slightly stronger than dwarfs and significantly weaker than giants in their molecular absorption.
In the bottom panels, the H$_2$O 1.74 $\mu$m spectral index in H band is plotted vs the CO 2.30 $\mu$m EW in K band (left) and the H$_2$O 1.34 $\mu$m spectral index in J band (right).
}
\end{figure*}

\begin{figure}
\centering
\includegraphics[width=\linewidth]{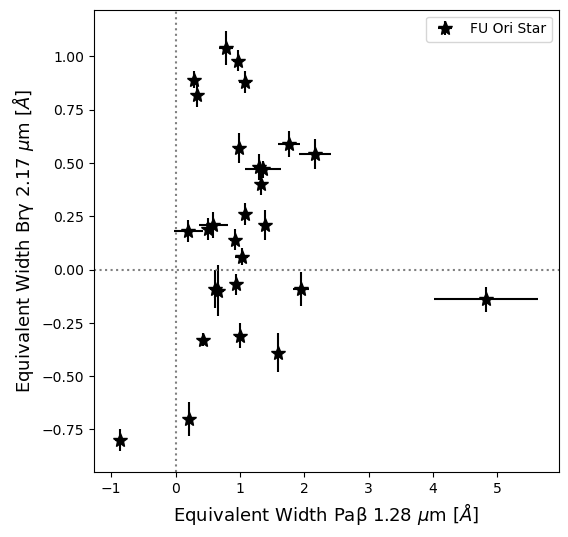}
\caption{\label{fig:wind}
Hydrogen line strengths in Pa$\beta$ 1.28 $\mu$m and Br$\gamma$ 2.17 $\mu$m EWs, illustrating wind-sensitive spectral features in the near-infrared. 
FUOrs generally have a strong Pa$\beta$ 1.28 $\mu$m absorption signature (typically blueshifted), 
but Br$\gamma$ 2.17 $\mu$m is less likely to exhibit absorption.
}
\end{figure}

\subsection{Comparison of FUOr and Reference Spectra}\label{sec:control}

Our goal is to identify which spectral features are distinct in FUOrs. 
As discussed in \S\ref{sec:irtf}, we use the IRTF Spectral Library to compare standard main sequence dwarfs and evolved giant and supergiant stars with FUOrs.  We also consider other more exotic S-type and C-type stars exhibiting strong molecular band absorption that could potentially be similar to FUOrs. 
In this subsection, we describe our parsing of the IRTF spectral library.

First, the entire IRTF spectral library can be used to illustrate the sensitivity to temperature and surface gravity of the different features we have measured. 
In general, molecular features such as H$_2$O, FeH, and CO are significantly stronger in giants than dwarfs, with CO further sensitive to supergiant vs giant surface gravity.  
Many atomic features are temperature sensitive, with monotonic increases or decreases with temperature that can reverse over the FGKM spectral type range. 
Some atomic features are also surface-gravity sensitive, at least in particular temperature ranges. 
Certain lines of \ion{Al}{1}, \ion{K}{1}, \ion{Mg}{1}, \ion{Si}{1}, and \ion{Sr}{2}
clearly separate dwarfs and giants, with
\ion{Si}{1} 1.59 $\mu$m further discriminating supergiant vs giant surface gravity
(see additional discussion in Appendix~\ref{sec:gravity}).  
The temperature and gravity behavior of different lines is discussed in the context of 
Figures \ref{fig:k_band}, \ref{fig:h_band}, and \ref{fig:j_band}.  

To distinguish FUOrs from other stellar objects, we first consider the overlap between FUOr spectral signatures and normal stars of different temperature. 
Although F, G, and K stars are broadly similar to FUOrs in the optical, they are too hot to have a resemblance in the near-infrared. Here, FUOrs are much closer to M-type stars. Despite this fact, the EWs of certain atomic features in FGK stars do overlap with the FUOr sequence.
So as to eliminate confusion, we do not consider the FGK spectral types from the IRTF spectral library in most of our analysis. 

We also remove the late-type M, as well as the L and T stars. These cool objects have many absorption features that are more extreme than those of FUOrs. 
While this fact makes them easier to distinguish from FUOrs by eye, 
it interferes with the ability to measure reliable equivalent widths, 
in practice contaminating the continuum around our selected features. 
We thus retain only stars of type M6 or earlier.

Finally, among the S and C stars in the IRTF spectral library, we present the analysis only for the S stars.  Upon visual examination of the C stars, we see CO features in K band as well as CO, FeH, \ion{Mg}{1}, and \ion{Si}{1} in H band, consistent with features seen in FUOrs.  The C types have other large features due to molecules at the red end of the H and Y bands, and numerous atomic features that make them look quite different from FUOrs.  As is true for the late M, L, and T spectral types, the C type features are so strong that they contaminate our feature continuum measurements.  Because of this, we do not show our measurements for the C types and keep only the S stars in conjunction with M types in our plots.

\subsection{Diagnostic Features For FUOrs}\label{sec:tests}

After taking measurements of all features as described in \S\ref{sec:ew}, we selected 12 
{of the most diagnostic features} to illustrate. Figures \ref{fig:k_band}-\ref{fig:wind}
{show the FUOrs in various parameter spaces relative to the comparison dwarfs, giants, and supergiants. 
We note that some measurements have positive values, rather than the expected absorption 
in metal lines and molecular bands. This is the result of contamination by other absorption 
in our fixed continuum bands at certain spectral types, or noisy continuum in certain FUOrs;
this fact does not influence our findings, however, since we are most interested in 
the segregation of the FUOrs in different parameter spaces.}

We present plots modeled after CO 2.30 $\mu$m vs \ion{Na}{1} 2.21 $\mu$m $+$ \ion{Ca}{1} 2.26 $\mu$m from \citet{Connelly2018}. To begin, we recreated this K-band plot. However, our testing suggests that CO 2.30 $\mu$m vs \ion{Na}{1} 2.21 $\mu$m is a better choice. Though the combination of two weak atoms is intended to further isolate FUOrs from populations with strong atmoic features, the noise of adding two EWs makes the test less effective. This motivated the creation of plots that compare the two atomic features: \ion{Na}{1} 2.21 $\mu$m vs \ion{Ca}{1} 2.26 $\mu$m. The results of these two K-band tests are shown in Figure \ref{fig:k_band}.

We mirror these plots in H and J bands. In H band, we plot FeH 1.62 $\mu$m vs \ion{Si}{1} 1.59 $\mu$m and \ion{Si}{1} 1.59 $\mu$m vs \ion{Mg}{1} 1.50 $\mu$m in Figure \ref{fig:h_band}. In J band, the two best atomic lines we find to go along with H$_2$O 1.34 $\mu$m are \ion{Si}{1} 1.32 $\mu$m and \ion{Al}{1} 1.31 $\mu$m in Figure \ref{fig:j_band}.

We do not present any diagnostic plots in the Y band,
since the measurements generally have large errors due to the deterioration of signal-to-noise towards the blue for our red sources 
Nevertheless, we note that molecular TiO vs atomic \ion{Sr}{2} appears to be a robust diagnostic of FUOrs with TiO being stronger than in all dwarfs, giants, and supergiants, and \ion{Sr}{2} being generally (but not always) stronger than in dwarfs, overlapping the giant/supergiant sequences. 

We also show several diagnostics that compare features across the different YJHK bands. The crossband comparisons of the molecular features are shown in Figure \ref{fig:crossmolecule}.


Figure \ref{fig:wind} shows the two hydrogen lines in our spectral range,
Pa$\beta$ 1.28 $\mu$m and Br$\gamma$ 2.17 $\mu$m.
These lines, along with \ion{He}{1} 1.08 $\mu$m, typically measure wind/outflow in FUOrs.
For hotter $T_{max}$ values, however, the hydrogen lines could have some disk contribution to the absorption.
{The net of disk absorption and superimposed wind emission$+$P Cyg absorption is complicated,
and when reduced to a mere equivalent width, could lead to ambiguous results.  In practice,}
the Pa$\beta$ 1.28 $\mu$m line is near-ubiquitously present in absorption, but Br$\gamma$ 2.17 
is not always obviously in emission or absorption, so the EW measurements can be mixed 
with many weakly positive EWs but also some negative (net emission) values  
{reported.  When small, these EWs are often spurious measurements.} 
The \ion{He}{1} 1.08 $\mu$m line is usually strong in FUOrs,
but its sometimes complicated line profile makes an integrated EW not as meaningful as the line shape.
Furthermore, as discussed above for the \ion{Sr}{2} 1.03 $\mu$m and TiO 1.05 $\mu$m features, SNR degrades
towards the blue, rendering \ion{He}{1} 1.08 $\mu$m difficult to detect and measure.
Echoing \citet{Connelly2018}, we reiterate that
an observed blueshifted absorption or P Cygni line profile in \ion{He}{1} 1.08 $\mu$m 
is a solid wind diagnostic.
More quantitatively, we also suggest Pa$\beta$ EW measurements.
for identifying wind signatures in FUOrs. 

Other measured features that appear in Table \ref{tab:all_features} but are not discussed above have been deemed not diagnostically relevant for distinguishing FUOrs from other stellar sources. Despite many of the measured features being strong (see Figure~\ref{fig:atlas}), the line and band strength measurements were not significantly different than the feature strengths exhibited by the M dwarfs and giants.  

\section{Summary and Discussion}\label{sec:discussion}

We have presented a set of spectroscopic diagnostics across the near-infrared YJHK bands that allows distinction of FU Ori disk spectra from those of normal stars.  
Similar to \citet{Connelly2018}, who promoted the use of a
CO 2.30 $\mu$m vs \ion{Na}{1} 2.21 $\mu$m $+$ \ion{Ca}{1} 2.26 $\mu$m parameter space, 
we identify ranges in various feature strength plots where FUOrs congregate as a population,
relatively isolated from M-type dwarf and giant stars having similar molecular spectra. 

Some atomic species, such as \ion{Na}{1}, \ion{Ca}{1}, and \ion{Mg}{1}, are weaker in FUOrs 
than in all or most late-type dwarfs and giants, 
while others such as \ion{Si}{1} have measured line strengths either stronger than all dwarfs and giants
(e.g. 1.32 $\mu$m) or between those of dwarfs and giants (e.g. 1.59 $\mu$m).
Similarly for molecules, TiO and H$_2$O are generally stronger than in all dwarfs and giants.
CO and FeH, by contrast, have measured strengths for FUORs between those of dwarfs and giants. 

Considering the features present in different wavelength ranges, \ion{Sr}{2} 1.03 $\mu$m is typically present in Y band, as is molecular TiO 1.05 $\mu$m.
In J band, strong \ion{Si}{1} 1.32 $\mu$m is a very good indicator of FUOr status. We also expect to see a strong H$_2$O feature around 1.34 $\mu$m, as highlighted by \citet{Connelly2018}. In H band, we usually (but not always) see the presence of FeH 1.62 $\mu$m that is measured together with a CO $(\delta\nu=3)$ feature.  H$_2$O 1.74 $\mu$m is also strong in H band, again as previously highlighted by \citet{Connelly2018}. 
Atomic \ion{Si}{1} 1.59 $\mu$m versus \ion{Mg}{1} 1.50 $\mu$m is also diagnostic. 
In K band, CO 2.30 $\mu$m $(\delta\nu=2)$ is stronger than in dwarfs and weaker than in giants. We suggest CO 2.30 $\mu$m vs \ion{Na}{1} 2.21 $\mu$m as diagnostic, modifying the inclusion of \ion{Ca}{1} 2.26 $\mu$m by \citet{Connelly2018} as it adds scatter to the FUOr cloud. We also quantify the domain of FUOrs in the wind-diagnostic lines of Pa$\beta$ 1.28 $\mu$m and Br$\gamma$ 2.17 $\mu$m, with the former typically in strong absorption (along with \ion{He}{1} 1.08 $\mu$m) and the latter weak or absent.

Our spectroscopic diagnostics were measured from dereddened spectra, for which we selected 
a $T_{max} \approx 5900$ K disk model as an SED template.  Because the $T_{max}$ values will vary 
among FUOrs by up to few thousand K, this template is not ideal for every object, 
but it is suitable in the absence of independent and detailed modeling that determines $T_{max}$, which often requires both a broadly measured SED and high dispersion spectroscopy 
for each source.  However, in practice,  errors in the derived $A_V$ values have negligible impact on spectral line equivalent widths and minimal errors on spectral indices.

We have also presented revised $A_V$ estimates for all sources in our
northern FUOr sample.  These values are consistently lower extinction estimates than those 
reported by \citet{Connelly2018}. The systematic difference is mainly due to our acknowledgment of the potential 
for excess flux in K band, which is present in many FUOrs but not in FU Ori itself
\citep[the dereddening template utilized by][]{Connelly2018}. 
We also present the median FUOr spectrum and show that it is remarkably similar to the spectrum of FU Ori itself.

The highly extincted nature of most FUOrs means that the highest SNR spectra are
usually obtained in K band, with SNR degrading towards the bluer end of the near-infrared.
The key signature of FUOr disks is their mix of atomic and molecular features, which arise due to
the $T(r)$ gradient in the viscously dominated inner part of the disk. Extremely low SNR
spectra may not be sensitive to narrow atomic features, however, with only the stronger molecules measurable.
We emphasize the importance of obtaining sufficiently high SNR spectra to conclusively detect atomic features at $J$ band, even for the reddest sources.

We anticipate our diagnostics as useful 
for vetting new FUOr candidates as they are discovered in time domain photometric surveys. {We propose the following diagnostics for future determination of FUOr status. In addition to an observed photometric outburst, spectroscopic signs of a high accretion-rate FUOr disk spectrum include:
\begin{itemize}
    \item strong J band and H band water vapor, exceeding the depth seen in dwarfs (and by implication both giants and superigants); 
    \item intermediate K band CO that overlaps with the strength in dwarfs and giants, but is weaker than in supergiants;
    \item H-band CO mixed with FeH at 1.62 $\mu$m that is often but not always present, and overlaps with dwarfs but is weaker than in giants and supergiants;
    \item weak or absent metal lines in K band, notably \ion{Na}{1}, \ion{Ca}{1}, and \ion{Mg}{1} that are weaker than in all dwarfs, giants, and supergiants;
    \item strong metallic \ion{Si}{1} at 1.32 $\mu$m and 1.59 $\mu$m that somewhat overlaps with dwarfs, but in 1.32 $\mu$m is stronger than giants and supergiants and in 1.59 $\mu$m is weaker than giants and supergiants ;
    \item weak but present (though not always) metallic \ion{Mg}{1} at 1.50 $\mu$m;
    \item weak but present (though not always) metallic \ion{Al}{1} at 1.31 $\mu$m;
    \item \ion{H}{1} Pa$\beta$ at 1.28 $\mu$m and \ion{He}{1} at 1.08 $\mu$m that are net absorption profiles but could have P Cygni shape in early FUOr stages.
\end{itemize}
}

In conclusion, we present a series of 11 diagnostics plots created from measuring 12 near-infrared spectral features. By plotting FUOr candidates on the same parameter spaces, we assess many sources at once to determine if match the rest of the population. After using these, we can look at the actual spectrum to check that we can actually see the features and confirm that the source is a FUOr.

\begin{acknowledgements}
We are grateful to the staff at Palomar Observatory for enabling the observations described in this paper.
We thank Michael Connelley for providing the spectral data published in \citet{Connelly2018} that we have re-analyzed for Figure~\ref{fig:av_comp}.  {We also thank Mike, Ellen Lee, and our paper referee for comments that helped us to improve our presentation.}
This work was supported in part by NASA grant \#80NSSC23K0655.
\end{acknowledgements}

\facility{Palomar:(TSPEC)}

\bibliography{main_arxiv}{}

@article{Gordon2024,
    author = {{Gordon}, K. ~D.},
    title = "{dust_extinction: Interstellar Dust Extinction Models}",
    journal = {Journal of Open Source Software},
    year = 2024,
    doi = {10.21105/joss.07023}
}

@article{Gordon2009,
    author = {{Gordon}, K. ~D. and {Cartledge}, S. and {Clayton}, G. ~S.},
    title = "{FUSE Measurements of Far-Ultraviolet Extinction. III. The Dependence of R(V) and Discrete Feature Limits From 75 Galactic Sightlines}",
    journal = {\apj},
    year = 2009,
    doi = {10.1088/0004-637X/705/2/1320}
}

@article{Connelly2018,
    author = {{Connelley}, M.~S. and {Reipurth}, B.},
    title = "{A Near-IR Spectroscopic Survey of FU Orionis Objects}",
    journal = {\apj},
    year = 2018,
    doi = {0.3847/1538-4357/aaba7b}
}

@article{Connelly2010,
    author = {{Connelley}, M.~S. and {Greene}, T.~P.},
    title = "{A Near-infrared Spectroscopic Survey of Class I Protostars}",
    journal = {\apj},
    year = 2010,
    doi = {10.1088/0004-6256/140/5/1214}
}

@article{Messineo2021,
    author = {{Messineo}, M. and {Figer}, D.~F. and {Kudritzki}, R.~P. and {Zhu}, Q. and {Menten}, K.~M. and {Ivanov}, V.~D. and {Chen} C.~H.~R},
    title = "{New Infrared Spectral Indices of Luminous Cold Stars: From Early K to M Types}",
    journal = {\apj},
    year = 2021,
    doi = {10.3847/1538-3881/ac116b}
}

@article{Slesnick2004,
    author = {{Slesnick}, C.~L. and {Hillenbrand}, L.~A. and {Carpenter}, J.~M.},
    title = "{The Spectroscopically Determined Substellar Mass Function of
the Orion Nebula Cluster}",
    journal = {\apj},
    year = 2004,
    doi = {10.1086/421898}
}

@article{Vollman2006,
    author = {{Vollman}, K. and {Eversberg}, T.},
    title = "{Remarks on statistical errors in equivalent widths}",
    journal = {Astronomische Nachrichten},
    year = 2006,
    doi = {10.1002/asna.2006}
}

@article{Sharon2010,
    author = {{Sharon}, C. and {Hillenbrand}, L. and {Fischer}, W. and {Edwards}, S.},
    title = "{F, G, K, M Spectral Standards in the Y Band (0.95-1.11 μm)}",
    journal = {\apj},
    year = 2010,
    doi = {10.1088/0004-6256/139/2/646}
}

@article{Hillenbrand2022,
    author = {{Hillenbrand}, L. ~A. and {Isaacson}, H. and {Rodriguez}, A. ~C. and {Connelley}, M. and {Reipurth}, B. and {Kuhn}, M. ~A. and {Beck}, T. and {Perez}, D. ~R.},
    title = "{LkHα 225 (V1318 Cyg) South in Outburst}",
    journal = {\apj},
    year = 2022,
    doi = {10.3847/1538-3881/ac4752}
}

@article{Herter2008,
    author = {{Herter}, T. ~L. and {Henderson}, C. ~P. and {Wilson}, J. ~C. and {Matthews}, K. ~Y. and {Rahmer}, G. and {Bonati}, M. and {Muirhead}, P. ~S. and {Adams}, J. ~D. and {Lloyd}, J. ~P. and {Skrutskie}, M. ~F. and {Moon}, D. and {Parshley}, S. ~C. and {Nelson}, M. ~J. and {Martinache}, F. and {Gull}, G. ~E.},
    title = "{The performance of TripleSpec at Palomar}",
    journal = {SPIE},
    year = 2008,
    doi = {10.1117/12.789660}
}

@article{Carvalho2024,
    author = {{Carvalho}, A. ~S. and {Hillenbrand}, L. ~A. and {France}, K. and {Herczeg}, G. ~J.},
    title = "{A Far-ultraviolet-detected Accretion Shock at the Star–Disk Boundary of FU Ori}",
    journal = {\apj},
    year = 2024,
    doi = {10.3847/2041-8213/ad74eb}
}

@ARTICLE{Allard2013,
       author = {{Allard}, F. and {Homeier}, D. and {Freytag}, B. and {Schaffenberger}, W. and {Rajpurohit}, A.~S.},
        title = "{Progress in modeling very low mass stars, brown dwarfs, and planetary mass objects.}",
      journal = {Memorie della Societa Astronomica Italiana Supplementi},
         year = 2013,
        month = jan,
       volume = {24},
        pages = {128},
          doi = {10.48550/arXiv.1302.6559},
archivePrefix = {arXiv},
       eprint = {1302.6559},
 primaryClass = {astro-ph.SR},
       adsurl = {https://ui.adsabs.harvard.edu/abs/2013MSAIS..24..128A}
}

@ARTICLE{Herbig1977,
       author = {{Herbig}, G.~H.},
        title = "{Eruptive phenomena in early stellar evolution.}",
      journal = {\apj},
         year = 1977,
        month = nov,
       volume = {217},
        pages = {693-715},
          doi = {10.1086/155615},
       adsurl = {https://ui.adsabs.harvard.edu/abs/1977ApJ...217..693H}
}

@incollection{protostars,
    author = {{Tsukamoto}, Y. and {Maury}, A. and {Commerçon}, B. and {Alves}, F. ~O. and {Cox}, E. ~G. and {Sakai}, N. and {Ray}, T. and {Zhao}, B. and {Machida}, M. ~N.},
    title = "{The Role of Magnetic Fields in the Formation of Protostars, Disks, and Outflows}",
    booktitle = "{Protostars and Planets VII}",
    publisher = {\pasp},
    year = 2023
}

@article{Le2024,
    author = {{Le Gouellec}, V. ~J. ~M. and {Greene}, T. ~P. and {Hillenbrand}, L. ~A. and {Yates}, Z.},
    title = "{New Insights on the Accretion Properties of Class 0 Protostars from 2 $\mu$m Spectroscopy}",
    journal = {\apj},
    year = 2024,
    doi = {10.3847/1538-4357/ad2935}
}

@article{Suresh2024,
    author = {{Suresh}, A. and {Karambelkar}, V. and {Kasliwal}, M. ~A. and {Ashley}, M. ~C. ~B. and {De}, K. and {Hankins}, M. ~J.},
    title = "{An Automated Catalog of Long Period Variables using Infrared Lightcurves from Palomar Gattini-IR}",
    journal = {\pasp},
    year = 2024,
    doi = {10.1088/1538-3873/ad68a4}
}

@article{Petrov2008,
    author = {{Petrov}, P. ~P. and {Herbig}, G. ~H.},
    title = "{Line Structure in the Spectrum of FU Orionis}",
    journal = {\aj},
    year = 2008,
    doi = {10.1088/0004-6256/136/2/676}
}

@article{Greene1996,
    author = {{Greene}, T. ~P. and {Lada}, C. ~J.},
    title = "{Near-Infrared Spectra and the Evolutionary Status of Young Stellar Objects: Results of a 1.1-2.4 $\mu$m Survey}",
    journal = {\aj},
    year = 1996,
    doi = {10.1086/118173}
}

@article{Kenyon1987,
    author = {{Kenyon}, S. ~J. and {Hartmann}, L.},
    title = "{Spectral Energy Distributions of T Tauri Stars: Disk Flaring and Limits on Accretion}",
    journal = {\apj},
    year = 1987,
    doi = {10.1086/165866}
}

@incollection{Fischer2023,
    author = {{Fischer}, W. ~J. and {Hillenbrand}, L. ~A. and {Herczeg}, G. ~J. and {Johnstone}, D. and {Kóspál}, Á. and {Dunham}, M. ~M.},
    title = "{Accretion Variability as a Guide to Stellar Mass Assembly}",
    booktitle = "Protostars and Planets VII",
    publisher = {\pasp},
    year = 2023
}

@article{Clarke2005,
    author = {{Clarke}, C. and {Lodato}, G. and {Melnikov}, S. ~Y. and {Ibrahimov}, M. ~A.},
    title = "{The photometric evolution of FU Orionis objects: disc instability and wind-envelope interaction}",
    journal = {\mnras},
    year = 2005,
    doi = {10.1111/j.1365-2966.2005.09231.x}
}

@article{Hartmann1996,
    author = {{Hartmann}, L. and {Kenyon}, S. ~J.},
    title = "{The FU Orionis Phenomenon}",
    journal = {\araa},
    year = 1996,
    doi = {10.1146/annurev.astro.34.1.207}
}

@article{Pena2019,
    author = {{Contreras Peña}, C. and {Naylor}, T. and {Morrell}, S.},
    title = "{Determining the recurrence time-scale of long-lasting YSO outbursts}",
    journal = {\mnras},
    year = 2019,
    doi = {10.1093/mnras/stz1019}
}

@article{Nagy2023,
    author = {{Nagy}, Z. and {Park}, S. and {Ábrahám}, P. and {Kóspál}, A. and {Cruz-Sáenz}, F.},
    title = "{The Gaia alerted fading of the FUor-type star Gaia21elv}",
    journal = {\mnras},
    year = 2023,
    doi = {10.1093/mnras/stad2019}
}

@article{Hillenbrand2023,
    author = {{Hillenbrand}, L. ~A. and {Carvalho}, A. and {van Roestel}, J. and {De}, K.},
    title = "{RNO 54: A Previously Unappreciated FU Ori Star}",
    journal = {\apjl},
    year = 2023,
    doi = {10.3847/2041-8213/ad0be0}
}

@article{Welty1992,
    author = {{Welty}, A. ~D. and {Strom}, S. ~E. and {Edwards}, S. and {Kenyon}, S. ~L. and {Hartmann}, L. ~W.},
    title = "{Optical Spectroscopy of Z Canis Majoris, V1057 Cygni, and FU Orionis: Accretion Disks and Signatures of Disk Winds}",
    journal = {\apj},
    year = 1992,
    doi = {10.1086/171785}
}

@article{Cushing2004,
    author = {{Cushing}, M. ~C. and {Vacca}, W. ~D. and {Rayner}, J. ~T},
    title = "{Spextool: A Spectral Extraction Package for SpeX, a 0.8-5.5 Micron Cross-Dispersed Spectrograph}",
    journal = {\pasp},
    year = 2004,
    doi =  {10.1086/382907}
}

@article{Vacca2003,
    author = {{Vacca}, W. ~D. and {Cushing}, M. ~C. and {Rayner}, J. ~T},
    title = "{A Method of Correcting Near-Infrared Spectra for Telluric Absorption}",
    journal = {\pasp},
    year = 2003,
    doi = {10.1086/346193}
}

@article{Rayner2009,
    author = {{Rayner}, J. ~T and {Cushing}, M. ~C. and {Vacca}, W. ~D.},
    title = "{The Infrared Telescope Facility (IRTF) Spectral Library: Cool Stars}",
    journal = {\apjs},
    year = 2009,
    doi = {10.1088/0067-0049/185/2/289}
}

@ARTICLE{Andrievsky2011,
       author = {{Andrievsky}, S.~M. and {Spite}, F. and {Korotin}, S.~A. and {Fran{\c{c}}ois}, P. and {Spite}, M. and {Bonifacio}, P. and {Cayrel}, R. and {Hill}, V.},
        title = "{NLTE strontium abundance in a sample of extremely metal poor stars and the Sr/Ba ratio in the early Galaxy}",
      journal = {\aap},
         year = 2011,
        month = jun,
       volume = {530},
          eid = {A105},
        pages = {A105},
          doi = {10.1051/0004-6361/201116591},
archivePrefix = {arXiv},
       eprint = {1104.0476},
 primaryClass = {astro-ph.SR},
       adsurl = {https://ui.adsabs.harvard.edu/abs/2011A&A...530A.105A}
}

@phdthesis{CarvalhoThesis2025,
    author = {{Carvalho}, A. ~S.},
    title = "{A Detailed Study of the Inner Disks of FU Ori Objects and the Star-Disk Boundary}",
    school = {California Institute of Technology},
    year = {2026},
    doi = {10.7907/4e53-3n79}
}

@ARTICLE{cushing2005,
       author = {{Cushing}, Michael C. and {Rayner}, John T. and {Vacca}, William D.},
        title = "{An Infrared Spectroscopic Sequence of M, L, and T Dwarfs}",
      journal = {\apj},
         year = 2005,
        month = apr,
       volume = {623},
       number = {2},
        pages = {1115-1140},
          doi = {10.1086/428040},
archivePrefix = {arXiv},
       eprint = {astro-ph/0412313},
 primaryClass = {astro-ph},
       adsurl = {https://ui.adsabs.harvard.edu/abs/2005ApJ...623.1115C}
}

@ARTICLE{Liu2022,
       author = {{Liu}, Hanpu and {Herczeg}, Gregory J. and {Johnstone}, Doug and {Contreras-Pe{\~n}a}, Carlos and {Lee}, Jeong-Eun and {Yang}, Haifeng and {Zhou}, Xingyu and {Yoon}, Sung-Yong and {Lee}, Ho-Gyu and {Kunitomo}, Masanobu and {Jose}, Jessy},
        title = "{Diagnosing FU Ori-like Sources: The Parameter Space of Viscously Heated Disks in the Optical and Near-infrared}",
      journal = {\apj},
         year = 2022,
        month = sep,
       volume = {936},
       number = {2},
          eid = {152},
        pages = {152},
          doi = {10.3847/1538-4357/ac84d2},
archivePrefix = {arXiv},
       eprint = {2207.13324},
 primaryClass = {astro-ph.SR},
       adsurl = {https://ui.adsabs.harvard.edu/abs/2022ApJ...936..152L}
}

@article{Andrae2010,
    author = {{Andrae}, R. and {Schulze-Hartung}, T. and {Melchior}, P.},
    title = "{Dos and don'ts of reduced chi-squared}",
    journal = {Instrumentation and Methods for Astrophysics},
    year = 2010,
    doi = {10.48550/arXiv.1012.3754} 
}

@article{Siwak2025,
    author = {{Siwak}, M. and {Kóspál}, A. and {Ábrahám}, A. and {Marton}, G. and {Zielinski}, P. and {Gromadzk}, M. and {Wyrzykowski}, Ł},
    title = "{Gaia20bdk – New FU Ori-type star in the Sh 2-301 star-forming region}",
    journal = {\aap},
    year = 2025,
    doi = {10.1051/0004-6361/202451061}
}

@article{Szegedi2020,
    author = {{Szegedi-Elek}, E. and {Ábrahám}, P. and {Wyrzykowski}, L. and {Kun}, M. and {Kóspál}, Á and {Chen}, L.},
    title = "{Gaia 18dvy: A New FUor in the Cygnus OB3 Association}",
    journal = {\apj},
    year = 2020,
    doi = {10.3847/1538-4357/aba129}
}

@article{Hillenbrand2021,
    author = {{Hillenbrand}, L. ~A. and {De}, K. and {Hankins}, M. and {Kasliwal}, M, ~M. and {Rebull}, L. ~M. and {Lau}, R. ~M.},
    title = "{Outbursting Young Stellar Object PGIR 20dci in the Perseus Arm}",
    journal = {\aj},
    year = 2021,
    doi = {10.3847/1538-3881/abe406}
}

@article{Hillenbrand2018,
    author = {{Hillenbrand}, L. ~A. and {Contreras Peña}, C. and {Morrell}, S. and {Naylor}, T. and {Kuhn}, M. ~A},
    title = "{Gaia 17bpi: An FU Ori–type Outburst}",
    journal = {\apj},
    year = 2018,
    doi = {10.3847/1538-4357/aaf414}
}
\bibliographystyle{aasjournal}

\appendix{}

\section{Feature Strength Results}\label{sec:datatable}

Tables \ref{tab:datatable1} and \ref{tab:datatable2} show the full table of results on spectral feature strength, derived for each FUOr in our sample using the methods described in \S\ref{sec:methods}. {The first line refers to measurements taken on the median FUOr derived in \S\ref{sec:median}.}  We include information about data quality (SNR), the extinction fit ($A_V$), and the measurements for each feature. {The SNR is calculated by taking the mean flux divided by error in each band.}

Three of our sources had multiple observations, and in these cases, we combined the spectra weighting by signal-to-noise as follows:
\begin{equation}
    F=\frac{\sum F_n\times\sigma_n^{-2}}{\sum \sigma_n^{-2}}
\end{equation}
and
\begin{equation}
    \sigma=\sqrt{\sum \sigma_n^2}
\end{equation}
where $F$ and $\sigma$ are the flux and the error of the combined spectrum, and $F_n$ and $\sigma_n$ are the flux and the error of the $n$th observation. These three sources are denoted in Tables \ref{tab:datatable1} and \ref{tab:datatable2} by an asterisk ($^*$).

\begin{table}
\begin{flushleft}
\caption{Full Datatable}
\label{tab:datatable1}
\begin{tabular}{cccccccccc}
\toprule
Object & SNR (Y) & SNR (J) & SNR (H) & SNR (K) & $A_V$ & Pa$\beta$ 1.28 $\mu$m & \ion{Al}{1} 1.31 $\mu$m & \ion{Si}{1} 1.32 $\mu$m & H$_2$O 1.34 $\mu$m \\
       &         &         &         &           & [mag] &  [\AA]                &  [\AA]                 &  [\AA]            &                    \\ 
\midrule
Median FUOr & \nodata & \nodata & \nodata & \nodata & \nodata & $0.8\pm0.06$ & $0.60\pm0.05$ & $0.83\pm0.04$ & $0.79\pm0.06$ \\
\midrule
Gaia 20bdk & 67 & 161 & 320 & 394 & $5.2^{+1.8}_{-1.5}$ & $1.77\pm0.17$ & $1.16\pm0.07$ & $1.93\pm0.07$ & $0.68\pm0.01$ \\
V883 Ori & 13 & 117 & 548 & 857 & $18.9^{+2.0}_{-1.8}$ & $0.92\pm0.07$ & $0.62\pm0.05$ & $0.83\pm0.04$ & $0.77\pm0.01$ \\
V582 Aur & 40 & 109 & 243 & 367 & $6.1^{+1.9}_{-1.7}$ & $1.09\pm0.06$ & $0.64\pm0.06$ & $1.4\pm0.05$ & $0.74\pm0.01$ \\
PP 13S & 0 & 8 & 125 & 481 & $26.7^{+3.8}_{-3.7}$ & $0.6\pm0.22$ & $2.27\pm0.15$ & $2.3\pm0.08$ & $0.74\pm0.08$ \\
V890 Aur & 114 & 209 & 339 & 373 & $1.6^{+1.0}_{-0.9}$ & $0.66\pm0.1$ & $0.66\pm0.08$ & $1.73\pm0.07$ & $0.72\pm0.0$ \\
RNO 54 & 91 & 200 & 382 & 525 & $2.8^{+1.1}_{-0.9}$ & $0.62\pm0.08$ & $1.05\pm0.06$ & $1.66\pm0.06$ & $0.74\pm0.0$ \\
IRAS 05450+0019 & 4 & 33 & 129 & 285 & $20.9^{+4.2}_{-3.2}$ & $1.0\pm0.08$ & $1.41\pm0.06$ & $1.88\pm0.05$ & $0.72\pm0.02$ \\
AR 6a & 21 & 135 & 455 & 604 & $13.5^{+2.5}_{-2.1}$ & $0.34\pm0.05$ & $0.58\pm0.04$ & $0.54\pm0.04$ & $0.95\pm0.01$ \\
BBW 76 & 60 & 108 & 168 & 182 & $0.1^{+1.0}_{-0.1}$ & $0.29\pm0.06$ & $0.29\pm0.05$ & $0.15\pm0.07$ & $1.01\pm0.01$ \\
V960 Mon & 160 & 280 & 446 & 528 & $1.3^{+0.7}_{-0.7}$ & $-0.87\pm0.07$ & $0.78\pm0.05$ & $1.43\pm0.05$ & $0.77\pm0.0$ \\
FU Ori & 177 & 288 & 437 & 459 & $0.7^{+0.7}_{-0.7}$ & $1.95\pm0.12$ & $0.77\pm0.07$ & $0.9\pm0.05$ & $0.77\pm0.0$ \\
V900 Mon & 54 & 156 & 359 & 504 & $9.2^{+1.5}_{-1.4}$ & $1.39\pm0.07$ & $1.11\pm0.06$ & $1.44\pm0.06$ & $0.75\pm0.01$ \\
V2775 Ori & 6 & 80 & 413 & 574 & $21.5^{+2.7}_{-2.3}$ & $0.43\pm0.06$ & $0.78\pm0.07$ & $1.12\pm0.04$ & $0.73\pm0.01$ \\
L1551 IRS5 & 11 & 73 & 311 & 577 & $16.9^{+3.9}_{-3.1}$ & $1.6\pm0.05$ & $0.96\pm0.1$ & $1.03\pm0.08$ & $0.76\pm0.01$ \\
\midrule
Gaia 17bpi & 12 & 33 & 71 & 36 & $2.6^{+1.4}_{-1.1}$ & $0.21\pm0.06$ & $0.6\pm0.08$ & $1.45\pm0.06$ & $0.73\pm0.03$ \\
V2495 Cyg & 0 & 6 & 65 & 144 & $22.4^{+4.8}_{-5.1}$ & $0.2\pm0.23$ & $-0.37\pm0.29$ & $1.06\pm0.09$ & $0.72\pm0.12$ \\
Par 21 & 56 & 126 & 210 & 184 & $2.3^{+0.9}_{-0.8}$ & $1.29\pm0.07$ & $0.34\pm0.08$ & $0.38\pm0.05$ & $0.81\pm0.01$ \\
V733 Cep & 28 & 123 & 336 & 456 & $8.6^{+2.0}_{-1.6}$ & $0.99\pm0.08$ & $0.35\pm0.07$ & $0.49\pm0.09$ & $0.84\pm0.01$ \\
V1515 Cyg & 90 & 195 & 319 & 353 & $2.1^{+0.7}_{-0.6}$ & $0.5\pm0.06$ & $0.66\pm0.07$ & $0.45\pm0.05$ & $0.82\pm0.01$ \\
Gaia 18dvy & 38 & 106 & 210 & 225 & $2.8^{+0.9}_{-0.7}$ & $0.97\pm0.08$ & $-0.03\pm0.1$ & $0.1\pm0.11$ & $0.87\pm0.01$ \\
V2494 Cyg & 16 & 68 & 189 & 275 & $8.2^{+1.7}_{-1.3}$ & $1.33\pm0.07$ & $0.65\pm0.07$ & $0.47\pm0.09$ & $0.86\pm0.02$ \\
HBC 722 & 114 & 241 & 372 & 401 & $2.6^{+0.7}_{-0.6}$ & $1.08\pm0.09$ & $0.62\pm0.07$ & $0.33\pm0.08$ & $0.84\pm0.01$ \\
HH 354 & 0 & 2 & 37 & 61 & $22.0^{+5.8}_{-5.3}$ & $4.79\pm0.8$ & $2.14\pm0.27$ & $1.36\pm0.18$ & $0.76\pm0.34$ \\
V1057 Cyg & 192 & 349 & 463 & 489 & $1.6^{+0.7}_{-0.6}$ & $1.36\pm0.28$ & $0.38\pm0.03$ & $0.92\pm0.05$ & $0.83\pm0.0$ \\
V1735 Cyg & 92 & 311 & 555 & 624 & $9.4^{+1.9}_{-1.5}$ & $0.79\pm0.11$ & $0.46\pm0.06$ & $0.36\pm0.08$ & $0.79\pm0.0$ \\
\midrule
RNO 1B & 12 & 63 & 190 & 272 & $10.7^{+1.9}_{-1.6}$ & $0.94\pm0.03$ & $1.28\pm0.07$ & $1.47\pm0.04$ & $0.74\pm0.01$ \\
RNO 1C & 10 & 73 & 264 & 337 & $13.8^{+1.9}_{-1.6}$ & $1.03\pm0.1$ & $0.65\pm0.05$ & $0.6\pm0.04$ & $0.91\pm0.01$ \\
PGIR20 dci & 1 & 5 & 38 & 47 & $13.5^{+4.0}_{-4.5}$ & $2.17\pm0.25$ & $0.67\pm0.06$ & $1.68\pm0.08$ & $0.71\pm0.14$ \\
\bottomrule
\end{tabular}
\tablecomments{First half of the full datatable showing spectral line and band strength results for all FUOrs in our sample. Each object has measured signal-to-noise in multiple bands, $A_V$ and $\chi^2$ derived from dereddening, and feature strengths in J band. Results are continued in Table \ref{tab:datatable2}. \textbf{The horizontal line at the center of the table denotes the transition from January to July observations. The last three objects are combinations of observations from both dates, as discussed in Appendix \ref{sec:datatable}.}}
\end{flushleft}
\end{table}
\begin{table}
\begin{flushleft}
\caption{Full Datatable}
\label{tab:datatable2}
\begin{tabular}{ccccccccc}
\toprule
Object & \ion{Mg}{1} 1.50 $\mu$m & \ion{Si}{1} 1.59 $\mu$m & CO/FeH 1.62 $\mu$m & H$_2$O 1.74 $\mu$m & Br$\gamma$ 2.17 $\mu$m & \ion{Na}{1} 2.21 $\mu$m & \ion{Ca}{1} 2.26 $\mu$m & CO 2.30 $\mu$m \\
       & [\AA]                   &     [\AA]               & [\AA]           &                     &   [\AA]                &  [\AA]                 &  [\AA]            &   [\AA]            \\ 
\midrule
Median FUOr & $1.89\pm0.04$ & $2.17\pm0.08$ & $3.81\pm0.08$ & $0.84\pm0.03$ & $0.28\pm0.06$ & $0.98\pm0.04$ & $2.83\pm0.06$ & $26.53\pm0.45$ \\
\midrule
Gaia 20bdk & $1.65\pm0.05$ & $1.92\pm0.07$ & $2.78\pm0.09$ & $0.81\pm0.0$ & $0.59\pm0.06$ & $0.77\pm0.06$ & $2.53\pm0.06$ & $28.7\pm0.8$ \\
V883 Ori & $1.39\pm0.03$ & $1.7\pm0.06$ & $3.68\pm0.07$ & $0.86\pm0.0$ & $0.14\pm0.05$ & $0.29\pm0.03$ & $1.76\pm0.04$ & $18.07\pm0.21$ \\
V582 Aur & $-0.04\pm0.05$ & $0.86\pm0.05$ & $3.82\pm0.07$ & $0.83\pm0.0$ & $0.26\pm0.05$ & $0.2\pm0.06$ & $1.61\pm0.05$ & $20.18\pm0.28$ \\
PP 13S & $1.87\pm0.12$ & $1.83\pm0.07$ & $1.62\pm0.07$ & $0.86\pm0.01$ & $0.21\pm0.06$ & $0.13\pm0.04$ & $0.94\pm0.04$ & $15.31\pm0.19$ \\
V890 Aur & $1.51\pm0.07$ & $1.98\pm0.08$ & $4.56\pm0.09$ & $0.84\pm0.0$ & $-0.1\pm0.12$ & $0.34\pm0.06$ & $2.03\pm0.06$ & $22.66\pm0.42$ \\
RNO 54 & $1.32\pm0.05$ & $2.09\pm0.07$ & $4.49\pm0.09$ & $0.84\pm0.0$ & $-0.09\pm0.09$ & $0.42\pm0.06$ & $2.55\pm0.06$ & $23.36\pm0.39$ \\
IRAS 05450+0019 & $1.91\pm0.07$ & $2.94\pm0.09$ & $3.96\pm0.11$ & $0.81\pm0.01$ & $-0.31\pm0.06$ & $1.08\pm0.05$ & $2.66\pm0.07$ & $29.74\pm0.54$ \\
AR 6a & $1.74\pm0.08$ & $2.78\pm0.08$ & $-0.29\pm0.05$ & $0.9\pm0.0$ & $0.82\pm0.06$ & $-0.01\pm0.05$ & $2.36\pm0.03$ & $10.04\pm0.09$ \\
BBW 76 & $0.79\pm0.06$ & $2.12\pm0.04$ & $-0.97\pm0.05$ & $0.92\pm0.01$ & $0.89\pm0.04$ & $-0.41\pm0.06$ & $0.99\pm0.06$ & $3.52\pm0.08$ \\
V960 Mon & $1.41\pm0.04$ & $2.37\pm0.06$ & $1.84\pm0.06$ & $0.85\pm0.0$ & $-0.8\pm0.05$ & $0.57\pm0.05$ & $1.69\pm0.05$ & $17.71\pm0.26$ \\
FU Ori & $2.63\pm0.07$ & $2.55\pm0.11$ & $3.77\pm0.08$ & $0.83\pm0.0$ & $-0.09\pm0.08$ & $0.92\pm0.05$ & $2.63\pm0.06$ & $27.68\pm0.5$ \\
V900 Mon & $2.7\pm0.06$ & $3.39\pm0.09$ & $3.48\pm0.09$ & $0.83\pm0.0$ & $0.21\pm0.07$ & $1.16\pm0.05$ & $2.42\pm0.07$ & $28.0\pm0.5$ \\
V2775 Ori & $1.72\pm0.06$ & $2.43\pm0.1$ & $5.63\pm0.12$ & $0.83\pm0.0$ & $-0.33\pm0.03$ & $0.81\pm0.05$ & $2.41\pm0.06$ & $27.5\pm0.46$ \\
L1551 IRS5 & $1.79\pm0.09$ & $2.48\pm0.1$ & $6.11\pm0.14$ & $0.84\pm0.0$ & $-0.39\pm0.09$ & $1.07\pm0.06$ & $2.97\pm0.07$ & $29.78\pm0.53$ \\
\midrule
Gaia 17bpi & $2.19\pm0.11$ & $2.02\pm0.09$ & $2.38\pm0.07$ & $0.84\pm0.02$ & $-0.7\pm0.08$ & $1.06\pm0.1$ & $2.53\pm0.08$ & $17.25\pm0.37$ \\
V2495 Cyg & $2.93\pm0.16$ & $2.74\pm0.11$ & $5.94\pm0.15$ & $0.82\pm0.02$ & $0.17\pm0.05$ & $1.65\pm0.07$ & $0.44\pm0.09$ & $34.94\pm0.77$ \\
Par 21 & $2.17\pm0.07$ & $3.02\pm0.11$ & $3.1\pm0.09$ & $0.82\pm0.01$ & $0.48\pm0.06$ & $0.97\pm0.06$ & $2.5\pm0.07$ & $21.72\pm0.53$ \\
V733 Cep & $2.07\pm0.09$ & $2.1\pm0.11$ & $5.16\pm0.1$ & $0.83\pm0.0$ & $0.57\pm0.07$ & $0.65\pm0.08$ & $2.1\pm0.08$ & $26.87\pm0.56$ \\
V1515 Cyg & $2.59\pm0.08$ & $2.3\pm0.11$ & $6.27\pm0.13$ & $0.87\pm0.0$ & $0.19\pm0.05$ & $1.34\pm0.06$ & $1.97\pm0.08$ & $23.15\pm0.48$ \\
Gaia 18dvy & $2.84\pm0.11$ & $2.77\pm0.11$ & $6.99\pm0.13$ & $0.87\pm0.01$ & $0.98\pm0.05$ & $1.19\pm0.07$ & $2.24\pm0.07$ & $28.5\pm0.58$ \\
V2494 Cyg & $2.09\pm0.08$ & $2.37\pm0.11$ & $6.0\pm0.12$ & $0.86\pm0.01$ & $0.4\pm0.05$ & $0.5\pm0.07$ & $1.68\pm0.08$ & $26.11\pm0.47$ \\
HBC 722 & $1.9\pm0.1$ & $2.89\pm0.1$ & $5.42\pm0.11$ & $0.86\pm0.0$ & $0.88\pm0.05$ & $0.82\pm0.07$ & $1.89\pm0.08$ & $34.84\pm0.74$ \\
HH 354 & $2.02\pm0.25$ & $1.73\pm0.13$ & $4.28\pm0.1$ & $0.82\pm0.03$ & $-0.15\pm0.06$ & $0.52\pm0.06$ & $1.01\pm0.08$ & $36.43\pm0.69$ \\
V1057 Cyg & $-0.07\pm0.08$ & $0.98\pm0.04$ & $-0.59\pm0.04$ & $0.88\pm0.0$ & $0.47\pm0.04$ & $-0.64\pm0.08$ & $-0.14\pm0.05$ & $16.46\pm0.3$ \\
V1735 Cyg & $1.43\pm0.11$ & $1.33\pm0.1$ & $5.22\pm0.09$ & $0.8\pm0.0$ & $1.04\pm0.08$ & $0.12\pm0.08$ & $1.43\pm0.08$ & $27.02\pm0.57$ \\
\midrule
RNO 1B & $2.43\pm0.06$ & $2.55\pm0.11$ & $5.39\pm0.11$ & $0.81\pm0.01$ & $-0.07\pm0.05$ & $1.22\pm0.06$ & $2.47\pm0.09$ & $29.39\pm0.69$ \\
RNO 1C & $2.72\pm0.08$ & $3.51\pm0.12$ & $2.6\pm0.08$ & $0.87\pm0.0$ & $0.06\pm0.04$ & $0.49\pm0.07$ & $3.18\pm0.06$ & $24.94\pm0.24$ \\
PGIR 20dci & $2.53\pm0.08$ & $2.88\pm0.11$ & $6.42\pm0.14$ & $0.82\pm0.03$ & $0.54\pm0.07$ & $1.51\pm0.06$ & $1.8\pm0.07$ & $30.32\pm0.62$ \\
\bottomrule
\end{tabular}
\tablecomments{Second half of the full datatable showing  spectral line and band strength results for all FUOrs in our sample. Each object has feature strengths in H and K bands. Results are continued from Table \ref{tab:datatable1}. \textbf{The horizontal line at the center of the table denotes the transition from January to July observations. The last three objects end are combinations of observations from both dates, as discussed in Appendix \ref{sec:datatable}.}}
\end{flushleft}
\end{table}

\section{Spectral Atlas}\label{sec:objects}

\begin{figure*}[h!]
\centering
\includegraphics[width=\textwidth]{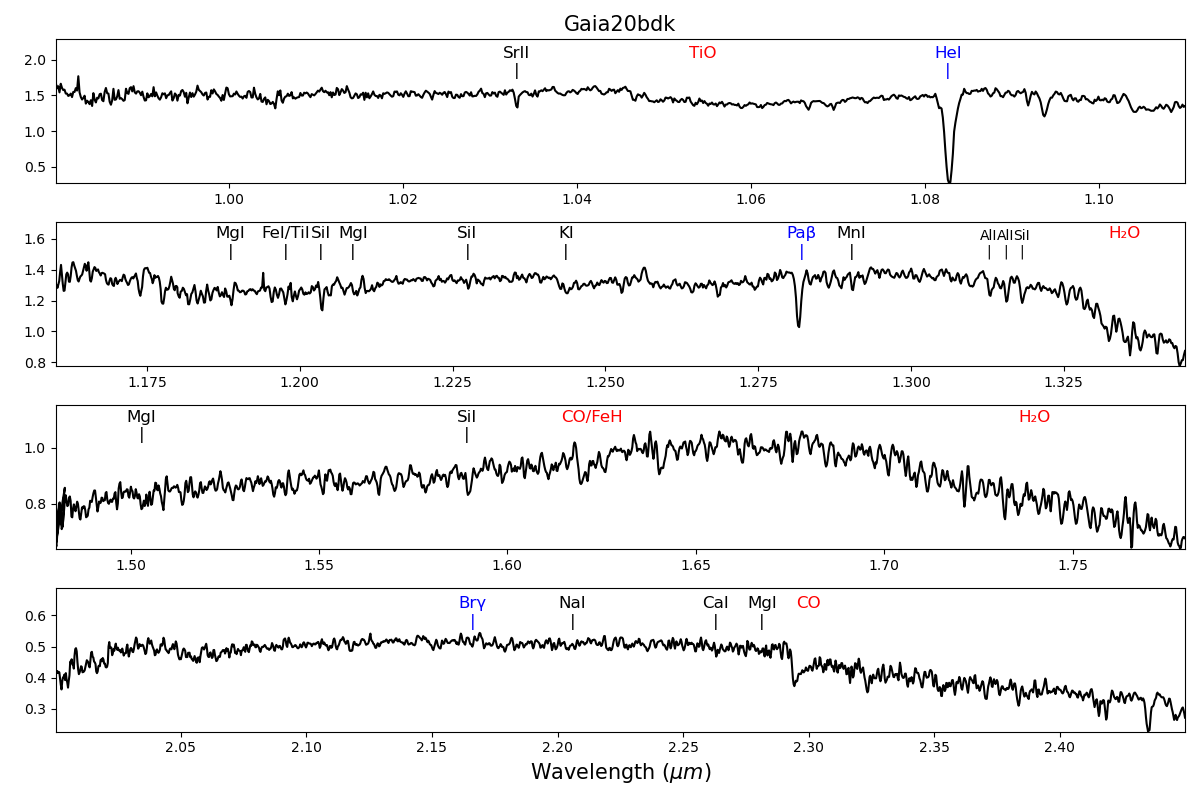}
\caption{
Example dereddened spectrum with all measured features labeled, for the case of Gaia 20bdk. The spectrum is normalized at 1.67 $\mu$m. The YJHK bands are separated into different panels, with Y band on top and K band on the bottom.
All measured features are labeled, with atomic features in black, molecular features in red, and outflow/wind feature in blue. 
{\it Figure is part of a Figure set available for each of the 28 sources in our sample.}
}
\label{fig:objects}
\end{figure*}

Figure \ref{fig:objects}, presented as a figure set, shows the full near-infrared spectrum for each object in our catalog, with labels indicating the spectral features that we measure. 

\newpage
\section{Measured Spectral Features} \label{sec:shading}

Figure \ref{fig:shaded} shows the defined measurement range and continuum region/s for each feature listed in Table \ref{tab:all_features}. The features are compared to the median spectrum from Figure \ref{fig:median}.


\begin{figure}[tbp]
\centering
\includegraphics[width=0.8\linewidth]{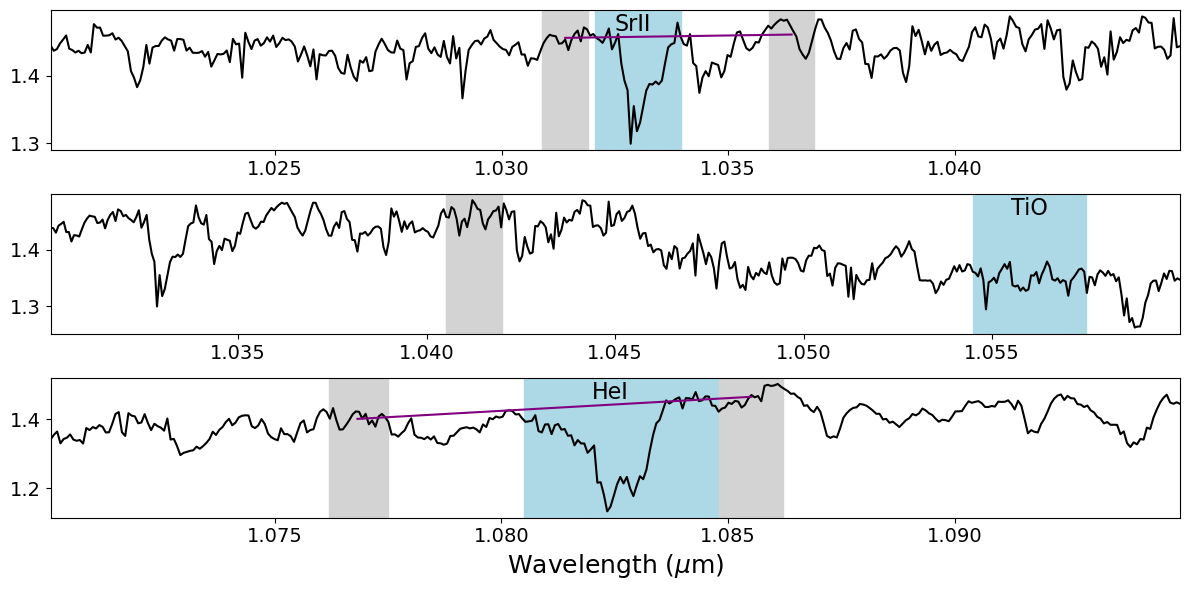}
\includegraphics[width=0.8\linewidth]{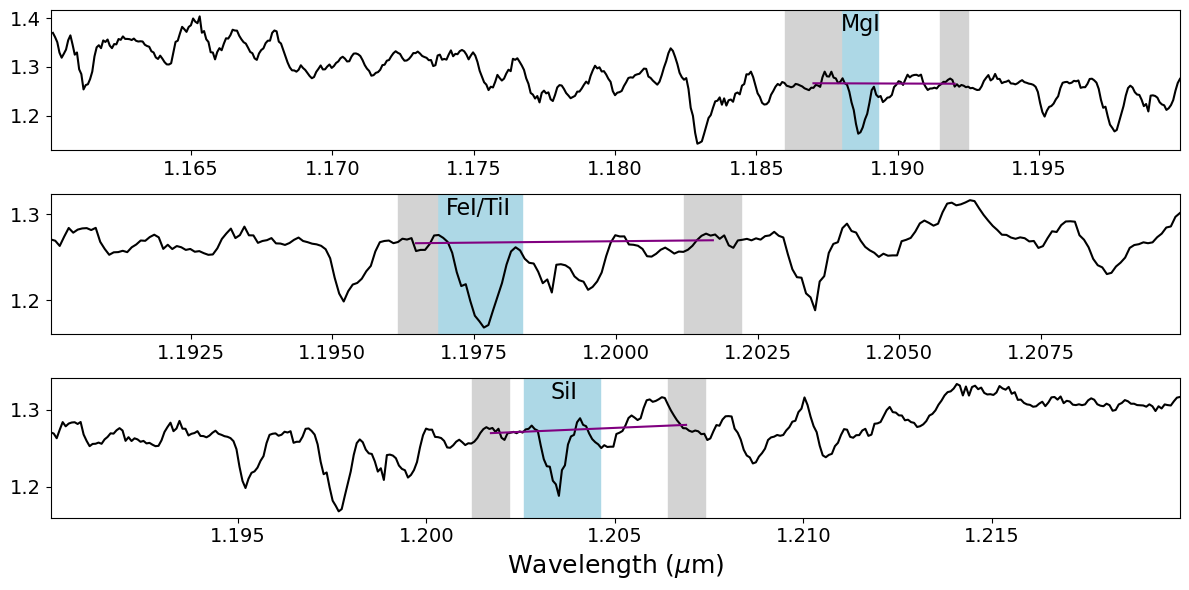}
\includegraphics[width=0.8\linewidth]{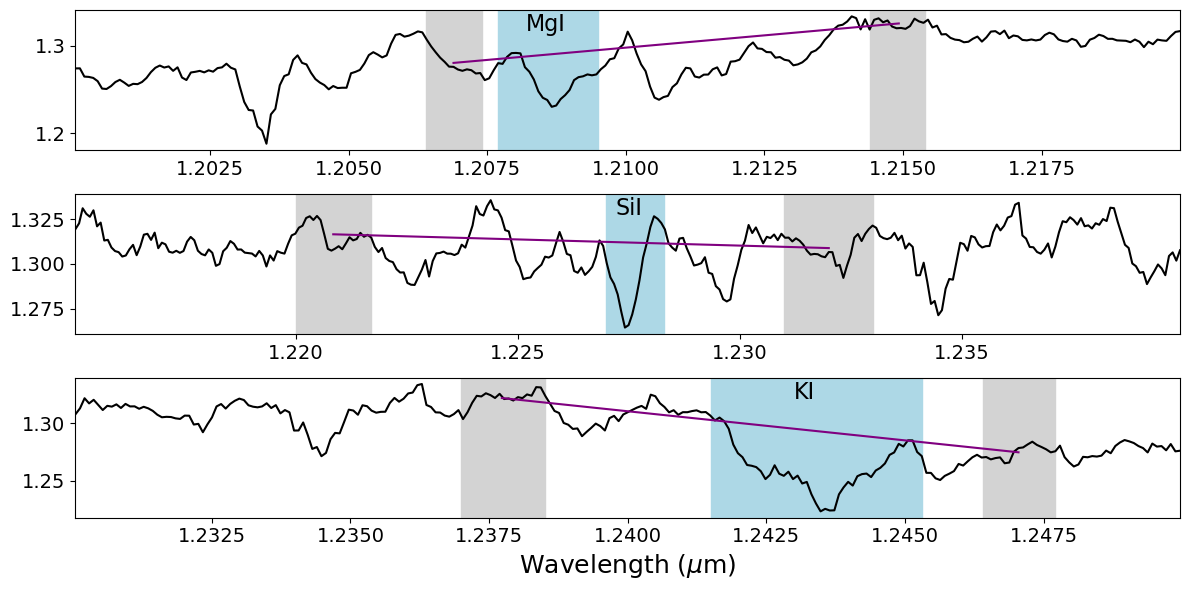}
\caption{}
\end{figure}

\begin{figure}[tbp]\ContinuedFloat
\centering
\includegraphics[width=0.8\linewidth]{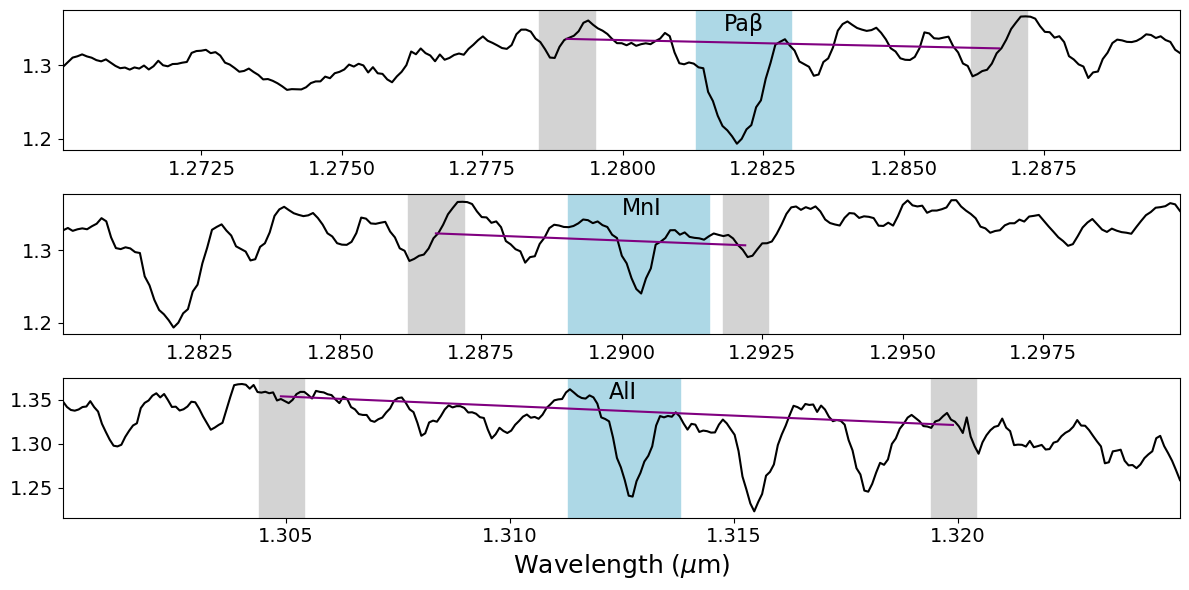}
\includegraphics[width=0.8\linewidth]{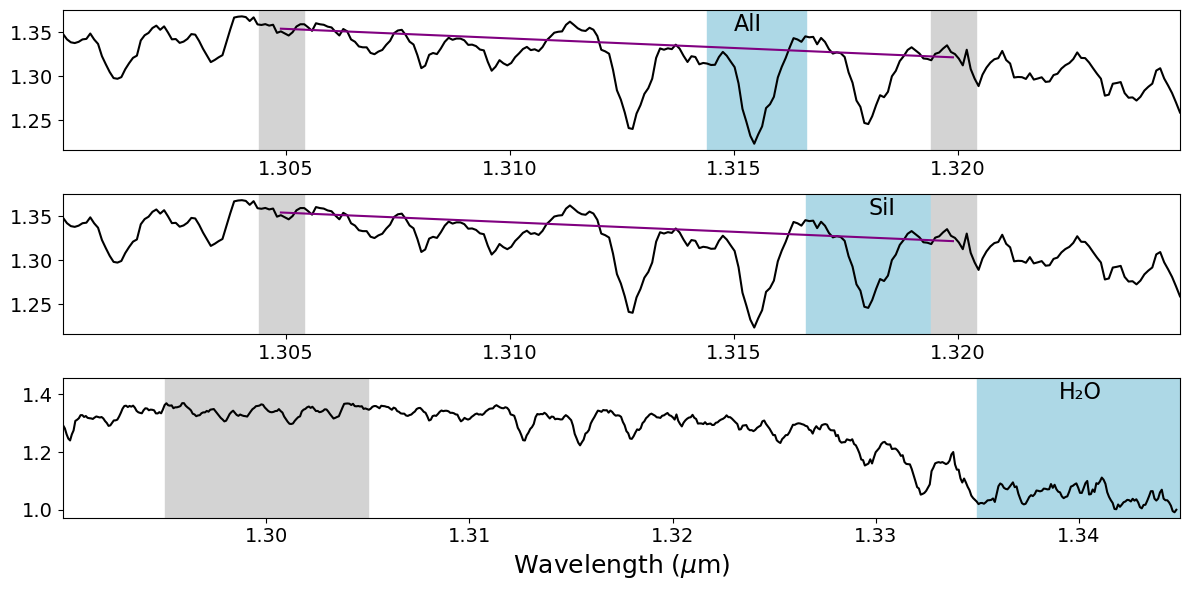}
\includegraphics[width=0.8\linewidth]{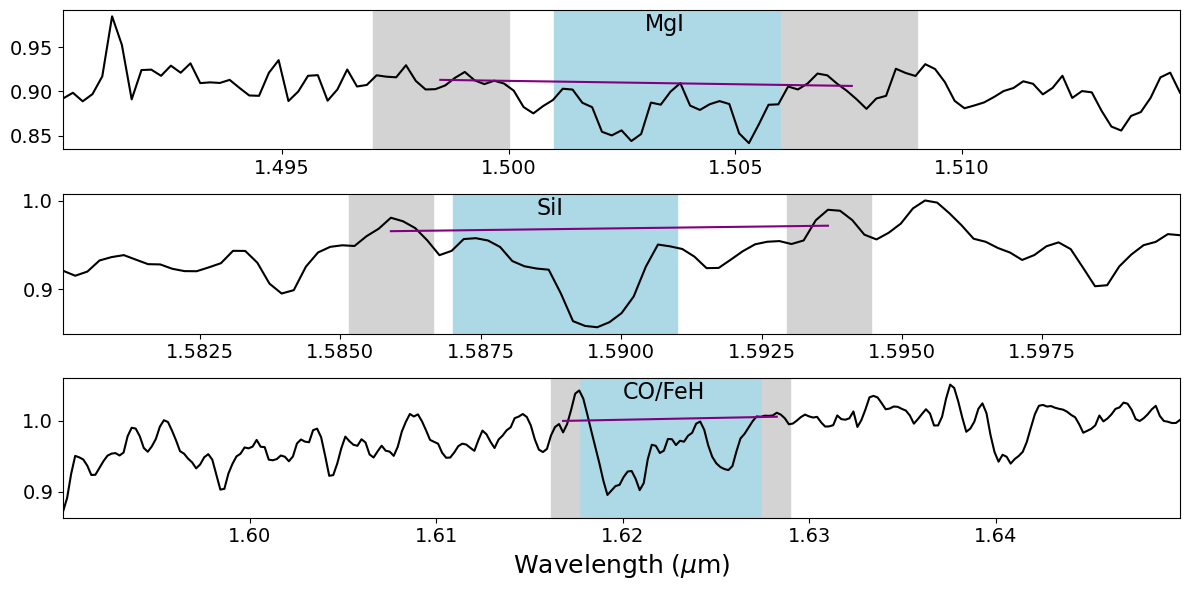}
\caption{}
\end{figure}

\begin{figure}[tbp]\ContinuedFloat
\centering
\includegraphics[width=0.8\linewidth]{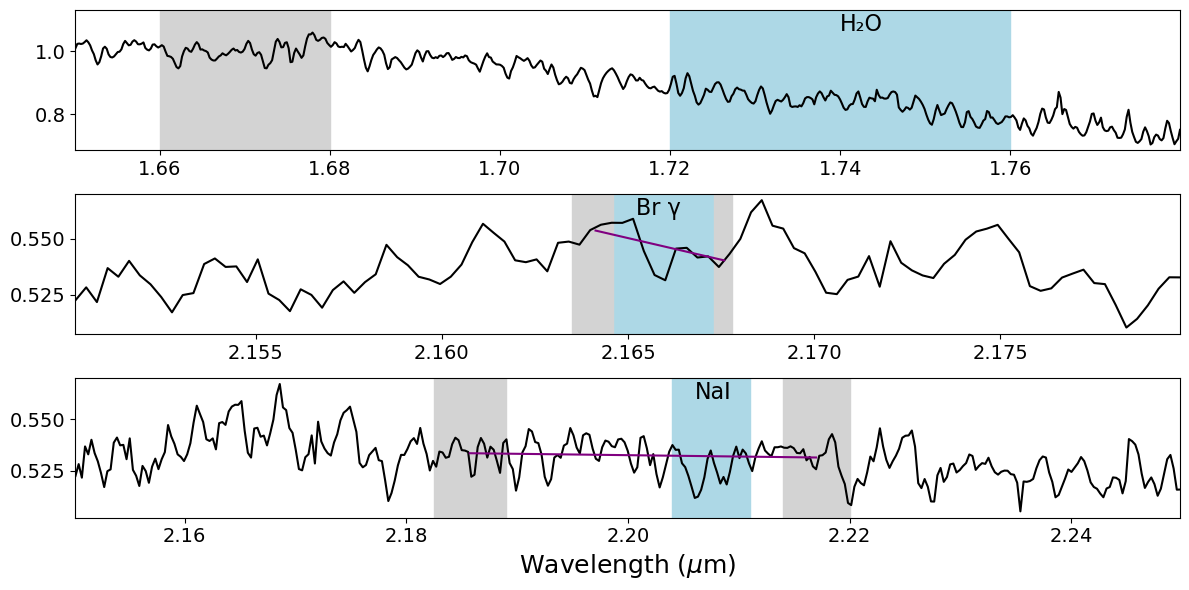}
\includegraphics[width=0.8\linewidth]{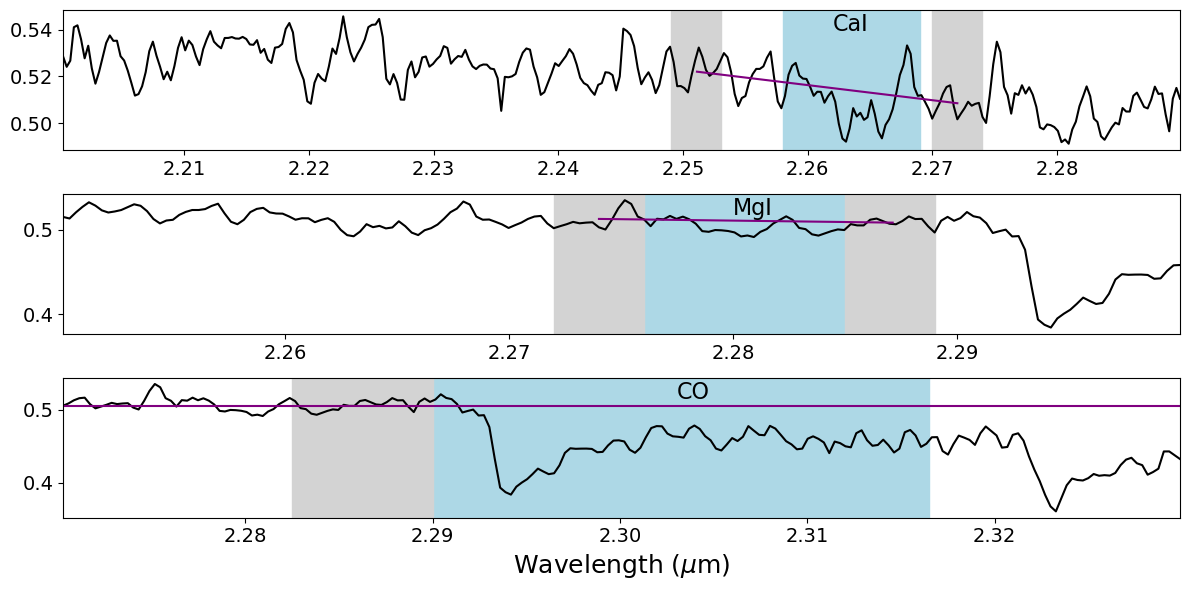}
\caption{
Full collection of measured features shown against the median FUOr spectrum from Figure \ref{fig:median}. The blue shaded area shows the measured region for each feature. The gray shaded area(s) show the nearby continuum region(s). The specific wavelength regions are defined in Table \ref{tab:all_features}.  Purple lines are the extrapolated continuum used for measuring the EWs.}
\label{fig:shaded}
\end{figure}

\section{Robust Atomic Surface Gravity Indicators}\label{sec:gravity}

\begin{figure}[h]
\centering
\includegraphics[width=0.5\linewidth]{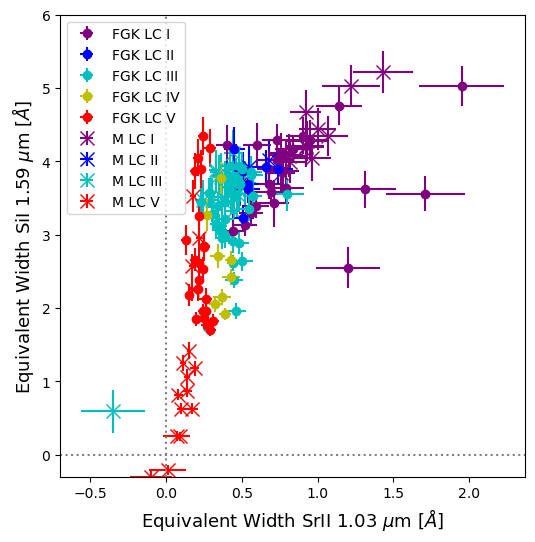}
\caption{
Equivalent widths of \ion{Sr}{2} 1.03 $\mu$m and \ion{Si}{1} 1.59 $\mu$m for stars in
the IRTF Spectral Library showing two features affected by surface gravity.  F, G, and K stars are denoted by dots while M stars are shown as crosses. Color represents luminosity class.}
\label{fig:surface_gravity}
\end{figure}

In Figure~\ref{fig:surface_gravity}, we illustrate the equivalent widths measured
for \ion{Sr}{2} 1.03 $\mu$m as a function of the \ion{Si}{1} 1.59 $\mu$m line strength.
The FGK and M stars are spread in temperature along the ordinate, with \ion{Si}{1} 1.59 $\mu$m strength
generally increasing towards hotter temperatures, though not monotonically at the hotter end
of the plotted range.  For the M types, the \ion{Si}{1} 1.59 $\mu$m strength is gravity-dependent,
as discussed in the main text, with the feature becoming stronger towards lower surface gravity.
For all spectral types, the I/II, III, and V luminosity classes clearly separate 
along the abscissa, with \ion{Sr}{2} 1.03 $\mu$m strength increasing towards lower surface gravity.

Ionized strontium (\ion{Sr}{2}) is a classic indicator of surface gravity in solar-type stars
with the most often discussed line in the optical, at 4078 \AA\ 
(along with its resonance line pair partner at 4215 \AA).
Additionally, there are two prominent \ion{Sr}{2} lines in the near-infrared Y band,
at 1.03 and 1.09 $\mu$m, that have been discussed as probes of surface gravity 
in normal stellar spectra \citep[e.g.][]{Sharon2010, Andrievsky2011}.
These lines have also been taken  as diagnostic of low gravity in FUOr disks \citep{Hillenbrand2022}. 

However, to date, there has been no clear empirical demonstration in the literature 
of the gravity sensitivity of the \ion{Sr}{2} lines in low-resolution spectra.
Figure~\ref{fig:surface_gravity} demonstrates the gravity sensitivity of \ion{Sr}{2} 1.03 $\mu$m  over a
wide range of temperature, and thus its diagnostic utility.

\end{document}